\documentclass[twocolumn,pre,aps,showpacs,showkeys,amsmath]{revtex4}
\usepackage{graphicx}

\begin{document}

\title{Cluster Burst Synchronization in A Scale-Free Network of Inhibitory Bursting Neurons}

\author{Sang-Yoon Kim}
\email{sykim@icn.re.kr}
\author{Woochang Lim}
\email{wclim@icn.re.kr}
\affiliation{Institute for Computational Neuroscience and Department of Science Education, Daegu National University of Education, Daegu 42411, Korea}

\begin{abstract}
We consider a scale-free network of inhibitory Hindmarsh-Rose (HR) bursting neurons, and investigate coupling-induced cluster burst synchronization by varying the average coupling strength $J_0$. For sufficiently small $J_0$, non-cluster desynchronized states exist. However, when passing a critical point $J^*_c~(\simeq 0.16)$, the whole population is segregated into 3 clusters via a constructive role of synaptic inhibition to stimulate dynamical clustering between individual burstings, and thus 3-cluster desynchronized states appear. As $J_0$ is further increased and passes a lower threshold $J^*_l~(\simeq 0.78)$, a transition to 3-cluster burst synchronization occurs due to another constructive role of synaptic inhibition to favor population synchronization. In this case, HR neurons in each cluster exhibit burst synchronization.
However, as $J_0$ passes an intermediate threshold $J^*_m~(\simeq 5.2)$, HR neurons begin to make intermittent hoppings between the 3 clusters. Due to the intermittent intercluster hoppings, the 3 clusters are integrated into a single one. In spite of break-up of the 3 clusters, (non-cluster) burst synchronization persists in the whole population, which is well visualized in the raster plot of burst onset times where bursting stripes (composed of  burst onset times and indicating burst synchronization) appear successively. With further increase in $J_0$, intercluster hoppings are intensified, and bursting stripes also become smeared more and more due to a destructive role of synaptic inhibition to spoil the burst synchronization. Eventually, when passing a higher threshold $J^*_h~(\simeq 17.8)$ a transition to desynchronization occurs via complete overlap between the bursting stripes. Finally, we also investigate the effects of stochastic noise on both 3-cluster burst synchronization and intercluster hoppings.
\end{abstract}


\maketitle

\section{Introduction}
\label{sec:INT}
Recently, much attention has been paid to burst synchronization in a population of bursting neurons \cite{PIR1,BSsync1,BSsync2,BSsync3,BSsync4,BSsync5,BSsync6,PIR2,BSsync7,BSsync8,BSsync9,BSsync10,BSsync11,BSsync12,BSsync13,BSsync14,BSsync15,BSsync16,BSsync17,BSsync18,BSsync19,BSsync20,BSsync21,PIR3,BSsync22,BSsync23,BSsync24,BSsync25,
BSsync26,Kim1,Kim2,NN-SFN,SBS}. Burstings occur when neuronal activity alternates, on a slow timescale, between a silent phase and an active (bursting) phase of fast repetitive spikings \cite{Burst2,Izhi,Burst1,Rinzel1,Rinzel2,Burst3}. Due to a repeated sequence of spikes in the bursting, there are several hypotheses on the importance of bursting activities in neural computation \cite{Burst2,Izhi2,Burst4,Burst5,Burst6}. For example, (a) bursts are necessary to overcome the synaptic transmission failure, (b) bursts are more reliable than single spikes in evoking responses in post-synaptic neurons, (c) bursts evoke long-term potentiation/depression (and hence affect synaptic plasticity much greater than single spikes), and (d) bursts can be used for selective communication between neurons. Intrinsically bursting neurons and chattering neurons in the cortex \cite{CT1,CT2}, thalamic relay neurons and thalamic reticular neurons in the thalamus \cite{TRN1,TRN2,TR}, hippocampal pyramidal neurons \cite{HP}, Purkinje cells in the cerebellum \cite{PC}, pancreatic $\beta$-cells \cite{PBC1,PBC2,PBC3}, and respiratory neurons in pre-Botzinger complex \cite{BC1,BC2} are representative examples of bursting neurons.

Here, we are concerned about burst synchronization (i.e., synchrony on the slow bursting timescale) which characterizes temporal coherence between burst onset times (i.e., times at which burstings start in active phases). This kind of burst synchronization is related to neural information processes in health and disease. For example, large-scale burst synchronization occurs in the sleep spindles through interaction between the excitatory thalamic relay cells and the inhibitory thalamic reticular neurons in the thalamus during the early stage of slow-wave sleep \cite{Spindle1,Spindle2}. These sleep spindles are associated with memory consolidation \cite{Spindle3,Spindle4}. In contrast, burst synchronization is also correlated to abnormal pathological rhythms, related to neural diseases such as movement disorder (Parkinson's disease and essential tremor) \cite{PD1,PD2,PD3,PD4,PD5} and epileptic seizure \cite{PD5,Epilepsy}.

In addition to burst synchronization, we are also interested in cluster synchronization. In this case, the whole population is segregated into synchronous sub-populations (called also as clusters) with phase lag among them \cite{CS1,CS2}. This type of cluster synchronization  has been investigated experimentally, numerically, or theoretically in a variety of contexts in diverse coupled (physical, chemical, biological, and neural) oscillators; to name a few, Josepson junction arrays \cite{Josep1,Josep2},  globally-coupled chemical oscillators \cite{CSExp1,CSExp2,CSExp3}, synthetic genetic networks \cite{Genetic}, and globally-coupled networks of inhibitory (non-oscillatory) reticular thalamic nucleus neurons \cite{PIR1} and other inhibitory model neurons \cite{PIR2,PIR3}.

Synaptic connectivity in neural networks has been found to have complex topology which is neither regular nor completely random \cite{Sporns,Buz2,CN1,CN2,CN3,CN4,CN5,CN6,CN7}.
Particularly, neural networks have been found to exhibit power-law degree distributions (i.e., scale-free property) in the rat hippocampal networks \citep{SF1,SF2,SF3,SF4} and the human cortical functional network \cite{SF5}. Moreover, robustness against simulated lesions of mammalian cortical anatomical networks \cite{SF6,SF7,SF8,SF9,SF10,SF11} has also been found to be most similar to that of a scale-free network (SFN) \citep{SF12}.
This type of SFNs are inhomogeneous ones with a few ``hubs'' (i.e., superconnected nodes) \cite{BA1,BA2}. Many recent works on various subjects of neurodynamics have been done in SFNs with a few percent of
hub neurons with an exceptionally large number of synapses \cite{BSsync11,BSsync12,BSsync14,BSsync15,BSsync19,BSsync26}.

In this paper, we consider an inhibitory SFN of suprathreshold (i.e., self-oscillating) Hindmarsh-Rose (HR) bursting neurons, and investigate coupling-induced cluster burst synchronization by changing the average coupling strength $J_0$.
For sufficiently small $J_0,$ desynchronized states exist. But, when passing a critical point $J^*_c~(\simeq 0.16)$, the whole population is segregated into 3 clusters via a constructive role of synaptic inhibition to stimulate dynamical clusterings between individual burstings, and thus 3-cluster desynchronized states appear.
In the presence of 3 clusters, inter-burst-intervals (IBIs) of individual HR neurons are localized in a region of $2~T_c < IBI < 4~T_c$ [$T_c:$ cluster period (i.e., time interval between appearance of successive clusters)], and a peak appears at $3~T_c$.
For $J_0 < J^*_c$, delocalization of IBIs occurs through crossing the left and/or the right boundaries (corresponding to $2~T_c$ and $4~T_c,$ respectively), and thus break-up of the 3 clusters occurs through intercluster hoppings between the clusters.

As $J_0$ is increased and a lower threshold $J^*_l~(\simeq 0.78)$ is passed, a transition to 3-cluster burst synchronization occurs due to another constructive role of synaptic inhibition to favor population synchronization.
In each cluster, HR neurons make burstings every 3rd cycle of the instantaneous whole-population burst rate $R_w(t)$ of the whole population, and hence a single peak appears at $3~T_G$ [$T_G:$ global period of $R_w(t)$] in the IBI histogram for the whole population of HR neurons. Moreover, these burstings in each cluster are also made in a coherent way, and hence a type of incomplete synchronization occurs in each cluster (i.e., burstings in each cluster show some coherence, although they are not completely synchronized). In this way, 3-cluster burst synchronization emerges.
This type of cluster burst synchronization is in contrast to that occurring via post-inhibitory rebound (PIR) in globally-coupled networks of subthreshold (i.e., non-oscillating) neurons with inhibitory synaptic connections \cite{PIR1,PIR2,PIR3}; in the case of PIR, complete synchronization appears in each cluster (i.e., states of all the neurons in each cluster are the same).

However, as $J_0$ is further increased and passes an intermediate threshold $J^*_m~(\simeq 5.2)$,
a new minor peak appears at $4~T_G$ in the IBI histogram, in addition to the major peak at $3~T_G$. Thus, delocalization of IBIs occurs by crossing the right boundary (corresponding to $4~T_G$). In this case, HR neurons intermittently fire burstings at a 4th cycle of $R_w(t)$ via burst skipping rather than at its 3rd cycle, and hence intermittent hoppings between the 3 clusters occur. Due to the intermittent intercluster hoppings via burst skippings, break-up of clusters occurs (i.e., the 3 clusters are integrated into a single one). However, in spite of break-up of the 3 clusters, burst synchronization persists in the whole population, which is well visualized in the raster plot of burst onset times where bursting stripes (composed of  burst onset times and representing burst synchronization) appear successively. With further increase in $J_0$, intercluster hoppings are intensified (e.g., for a larger $J_0$ a 3rd peak appears at $5~T_G$ in the IBI histogram), and bursting stripes also become smeared more and more due to a destructive role of synaptic inhibition to spoil the burst synchronization. Eventually, when passing a higher threshold $J^*_h~(\simeq 17.8),$ a transition to desynchronization occurs via complete overlap between the bursting stripes. In a desynchronized case, burst onset times are completely scattered without forming any stripes in the raster plot. Finally, the effects of stochastic noise on both 3-cluster burst synchronization and intercluster hoppings are also investigated.

This paper is organized as follows. In Sec.~\ref{sec:SFN}, we describe a Barab\'{a}si-Albert SFN composed of inhibitory HR bursting neurons. Then, in Sec.~\ref{sec:PS} we investigate coupling-induced cluster burst synchronization by varying the average coupling strength $J_0$, and then study the effects of stochastic noise on burst synchronization in Sec.~\ref{sec:Noise}. Finally, we give summary and discussion in Sec.~\ref{sec:SUM}.

\section{Inhibitory Scale-Free Network of Hindmarsh-Rose Bursting Neurons}
\label{sec:SFN}
We consider an inhibitory SFN composed of $N$ bursting neurons equidistantly placed on a one-dimensional ring of radius
${\frac {N} {2 \pi}}$. We employ a directed Barab\'{a}si-Albert SFN model (i.e. growth and preferential directed attachment) \citep{BA1,BA2}. At each discrete time $t,$ a new node is added, and it has $l_{in}$ incoming (afferent) edges and $l_{out}$ outgoing (efferent) edges via preferential attachments with $l_{in}$ (pre-existing) source nodes and $l_{out}$ (pre-existing) target nodes, respectively. The (pre-existing) source and target nodes $i$ (which are connected to the new node) are preferentially chosen depending on their out-degrees $d_i^{(out)}$ and in-degrees $d_i^{(in)}$ according to the attachment probabilities $\Pi_{source}(d_i^{(out)})$ and $\Pi_{target}(d_i^{(in)})$, respectively:
\begin{eqnarray}
\Pi_{source}(d_i^{(out)}) &=& \frac{d_i^{(out)}}{\sum_{j=1}^{N_{t -1}}d_j^{(out)}}\;\; \textrm{and} \nonumber \\
\Pi_{target}(d_i^{(in)}) &=& \frac{d_i^{(in)}}{\sum_{j=1}^{N_{t -1}}d_j^{(in)}},
\label{eq:AP}
\end{eqnarray}
where $N_{t-1}$ is the number of nodes at the time step $t-1$.
Here, we consider the case of symmetric preferential attachment with $l_{in} = l_{out} [\equiv l^* (=15)]$.
For generation of an SFN with $N$ nodes, we begin with the initial network at $t=0$, consisting of $N_0=50$ nodes where the node 1 is connected bidirectionally to all the other nodes, but the remaining nodes (except the node 1) are sparsely and randomly connected with a low probability $p=0.1$. The processes of growth and preferential attachment are repeated until the total number of nodes becomes $N$. In this case, the node 1 will be grown as the head hub with the highest degree.

\begin{table}
\caption{Parameter values used in our computations.}
\label{tab:Parm}
\begin{ruledtabular}
\begin{tabular}{llllll}
(1) & \multicolumn{5}{l}{Single HR Bursting Neurons \cite{Longtin}} \\
&  $a=1$ & $b=3$ & $c=1$ & $d=5$ & $r=0.001$ \\
&  $s=4$ & $x_0 = -1.6$ & & &  \\
\hline
(2) & \multicolumn{5}{l}{External Stimulus to HR Bursting Neurons} \\
& \multicolumn{2}{l}{$I_{DC,i} \in [1.3, 1.4]$} & \multicolumn{3}{l}{$D$: Varying} \\
\hline
(3) & \multicolumn{5}{l}{Inhibitory Synapse Mediated by The GABA$_{\rm A}$} \\
& \multicolumn{5}{l}{Neurotransmitter \cite{GABA}} \\
& $\tau_l=1$ & $\tau_r=0.5$ & $\tau_d=5$ & \multicolumn{2}{l}{$X_{syn}=-2$} \\
\hline
(4) & \multicolumn{5}{l}{Synaptic Connections between Neurons in The} \\
& \multicolumn{5}{l}{Barab\'{a}si-Albert SFN} \\
& \multicolumn{5}{l}{$l^*=15$ (symmetric preferential attachment)} \\
& $J_0:$ Varying & $\sigma_0 = 0.1$ \\
\end{tabular}
\end{ruledtabular}
\end{table}

As an element in our SFN, we choose the representative bursting HR neuron model which was originally introduced to describe the time evolution of the membrane potential for the pond snails \cite{HR1,HR2,HR3}.
We consider the Barab\'{a}si-Albert SFN composed of $N$ HR bursting neurons. The following equations (\ref{eq:PD1})-(\ref{eq:PD3}) govern the population dynamics in the SFN:
\begin{eqnarray}
\frac{dx_i}{dt} &=& y_i - a x^{3}_{i} + b x^{2}_{i} - z_i +I_{DC,i} +D \xi_{i} -I_{syn,i}, \label{eq:PD1} \\
\frac{dy_i}{dt} &=& c - d x^{2}_{i} - y_i, \label{eq:PD2} \\
\frac{dz_i}{dt} &=& r \left[ s (x_i - x_o) - z_i \right], \label{eq:PD3}
\end{eqnarray}
where
\begin{eqnarray}
I_{syn,i} &=& \frac{1}{d_i^{(in)}} \sum_{j=1 (j \ne i)}^N J_{ij} w_{ij} g_j(t) (x_i - X_{syn}), \label{eq:PD4}\\
g_j(t) &=& \sum_{f=1}^{F_j} E(t-t_f^{(j)}-\tau_l); \nonumber \\
E(t) &=& \frac{1}{\tau_d - \tau_r} (e^{-t/\tau_d} - e^{-t/\tau_r}) \Theta(t). \label{eq:PD5}
\end{eqnarray}
Here, the state of the $i$th neuron at a time $t$ (measured in units of milliseconds) is characterized by three state variables: the fast membrane potential $x_i$, the fast recovery current $y_i,$ and the slow adaptation current $z_i$. The parameter values used in our computations are listed in Table \ref{tab:Parm}. More details on external stimulus to each HR neuron, synaptic currents, and numerical integration of the governing equations are given in the following subsections.

\begin{figure}
\includegraphics[width=0.6\columnwidth]{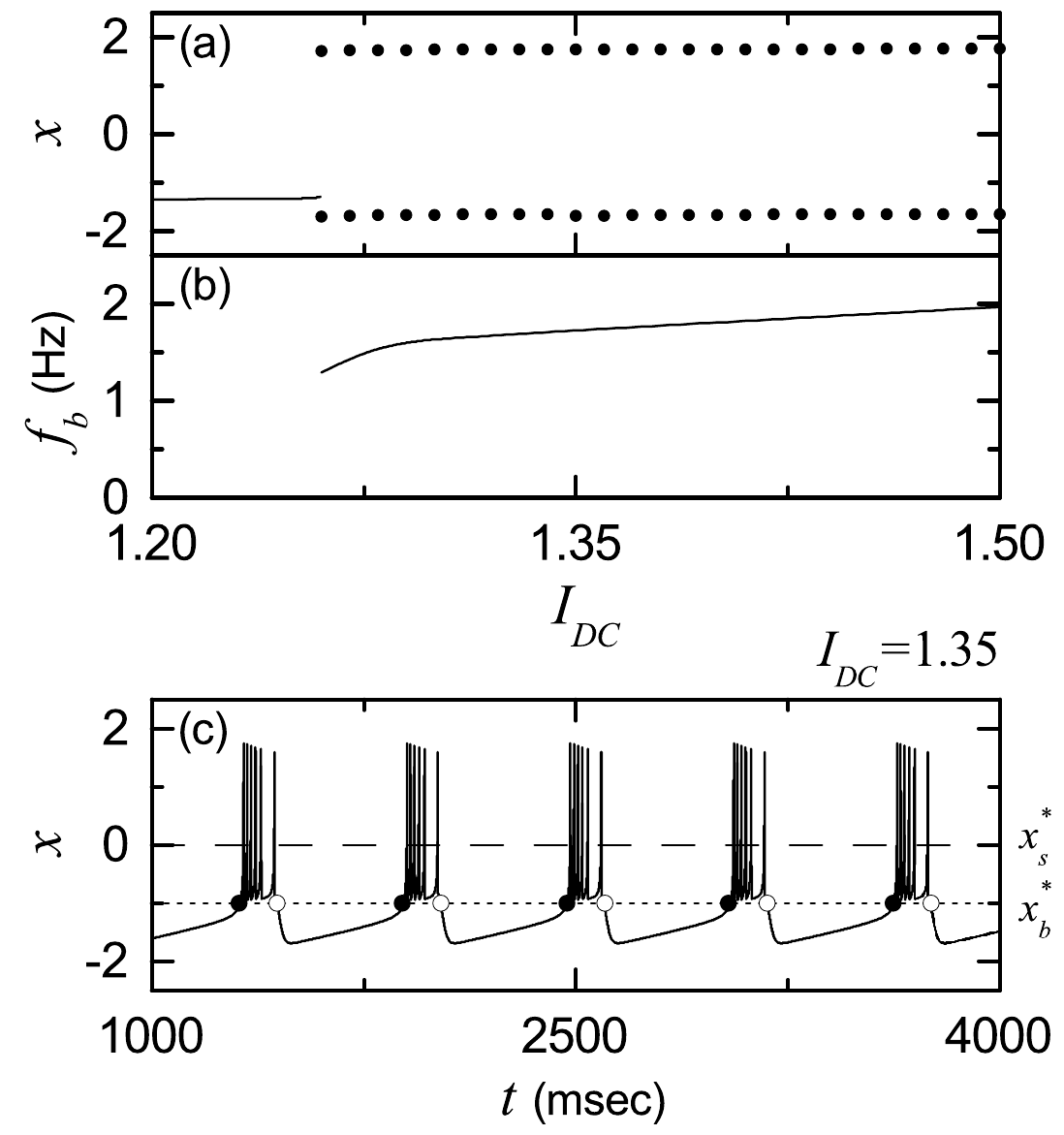}
\caption{
Single bursting HR neuron for $D=0$. (a) Bifurcation diagram in the single HR neuron. Solid line represents a stable resting state, while for the bursting state, maximum and minimum values of the membrane potential $x$ are denoted by solid circles.
(b) Bursting frequency $f_b$ versus $I_{DC}$.
(c) Time series of $x(t)$ for the bursting state when $I_{DC}=1.35$. The dotted horizontal line ($x^*_b=-1$) and the dashed horizontal line ($x^*_s=0$) denote the bursting and the spiking thresholds, respectively. The solid and open circles represent the burst onset and offset times, respectively.
}
\label{fig:Single}
\end{figure}

\subsection{External Stimulus to Each HR Neuron}
\label{subsubsec:Sti}
Each bursting HR neuron (whose parameter values are in the 1st item of Table \ref{tab:Parm} \cite{Longtin}) is stimulated by a DC current $I_{DC,i}$ and an independent Gaussian white noise $\xi_i$ [see the 5th and the 6th terms in Eq.~(\ref{eq:PD1})] satisfying $\langle \xi_i(t) \rangle =0$ and $\langle \xi_i(t)~\xi_j(t') \rangle = \delta_{ij}~\delta(t-t')$, where $\langle\cdots\rangle$ denotes the ensemble average. The intensity of noise $\xi_i$ is controlled by the parameter $D$.
As $I_{DC}$ passes a threshold $I_{DC}^* (\simeq 1.26)$ in the absence of noise (i.e., $D=0$), each single HR neuron exhibits a transition from a resting state to a bursting state [see Fig.~\ref{fig:Single}(a)]. With increasing $I_{DC}$, the bursting frequency $f_b$, (corresponding to the reciprocal of the average IBI $\langle IBI \rangle$), increases monotonically, as shown in Fig.~\ref{fig:Single}(b). For a suprathreshold case of $I_{DC}=1.35$, deterministic bursting occurs when neuronal activity alternates, on a slow time scale $(\simeq 578$ msec), between a silent phase and an active (bursting) phase of fast repetitive spikings, as shown in Fig.~\ref{fig:Single}(c). The dotted horizontal line ($x^*_b=-1$) denotes the bursting threshold (the solid and open circles denote the active phase onset and offset times, respectively), while the dashed horizontal line ($x^*_s=0$) represents the spiking threshold within the active phase. An active phase of the bursting activity begins (ends) at a burst onset (offset) time when the membrane potential $x$ of the bursting HR neuron passes the bursting threshold of
$x^*_b=-1$ from below (above). In this case, the HR neuron exhibits bursting activity with the slow bursting frequency $f_b~ (\simeq 1.7$ Hz) [corresponding to the reciprocal of the average IBI ($\langle IBI \rangle \simeq 578$ msec)]. Throughout this paper, we consider a suprathreshold case such that the value of $I_{DC,i}$ is chosen via uniform random sampling in the range of [1.3,1.4], as shown in the 2nd item of Table \ref{tab:Parm}.

\subsection{Synaptic Currents}
\label{subsec:Syn}
The last term in Eq.~(\ref{eq:PD1}) represents the synaptic couplings of HR bursting neurons.
The coupling strength of the synapse from the $j$th pre-synaptic neuron to the $i$th post-synaptic neuron is $J_{ij}$.
These synaptic strengths are normally distributed with the mean $J_0$ and the standard deviation $\sigma_0~(=0.1)$.
$I_{syn,i}$ of Eq.~(\ref{eq:PD4}) represents a synaptic current injected into the $i$th neuron, and $X_{syn}$ is the synaptic reversal potential. The synaptic connectivity is given by the connection weight matrix $W$ (=$\{ w_{ij} \}$) where  $w_{ij}=1$ if the bursting neuron $j$ is presynaptic to the bursting neuron $i$; otherwise, $w_{ij}=0$.
Here, the synaptic connection is modeled in terms of the Barab\'{a}si-Albert SFN. Then, the in-degree of the $i$th neuron, $d_i^{(in)}$ (i.e., the number of synaptic inputs to the neuron $i$) is given by $d_i^{(in)} = \sum_{j=1 (j \ne i)}^N w_{ij}$. The fraction of open synaptic ion channels at time $t$ is denoted by $g(t)$. The time course of $g_j(t)$ of the $j$th neuron is given by a sum of delayed double-exponential functions $E(t-t_f^{(j)}-\tau_l)$ [see Eq.~(\ref{eq:PD5})], where $\tau_l$ is the synaptic delay, and $t_f^{(j)}$ and $F_j$ are the $f$th spike and the total number of spikes of the $j$th neuron at time $t$, respectively. Here, $E(t)$ [which corresponds to contribution of a presynaptic spike occurring at time $0$ to $g_j(t)$ in the absence of synaptic delay] is controlled by the two synaptic time constants: synaptic rise time $\tau_r$ and decay time $\tau_d$, and $\Theta(t)$ is the Heaviside step function: $\Theta(t)=1$ for $t \geq 0$ and 0 for $t <0$. For the inhibitory GABAergic synapse (involving the $\rm{GABA_A}$ receptors), the values of $\tau_l$, $\tau_r$, $\tau_d$, and $X_{syn}$ are listed in the 3rd item of Table \ref{tab:Parm} \cite{GABA}.

\subsection{Numerical Integration}
\label{subsec:NI}
Numerical integration of differential equations (\ref{eq:PD1})-(\ref{eq:PD3}) is done by using the 4th-order Runge-Kutta  method in the absence of noise ($D=0$) and the Heun method \cite{SDE} in the presence of noise ($D>0$) (with the time step $\Delta t=0.01$ msec). For each realization, we choose a random initial point $[x_i(0),y_i(0),z_i(0)]$ for the $i$th $(i=1,\dots, N)$ neuron with uniform probability in the range of $x_i(0) \in (-1.5,1.5)$, $y_i(0) \in (-10,0)$, and $z_i(0) \in (1.2,1.5)$.

\begin{figure}
\includegraphics[width=\columnwidth]{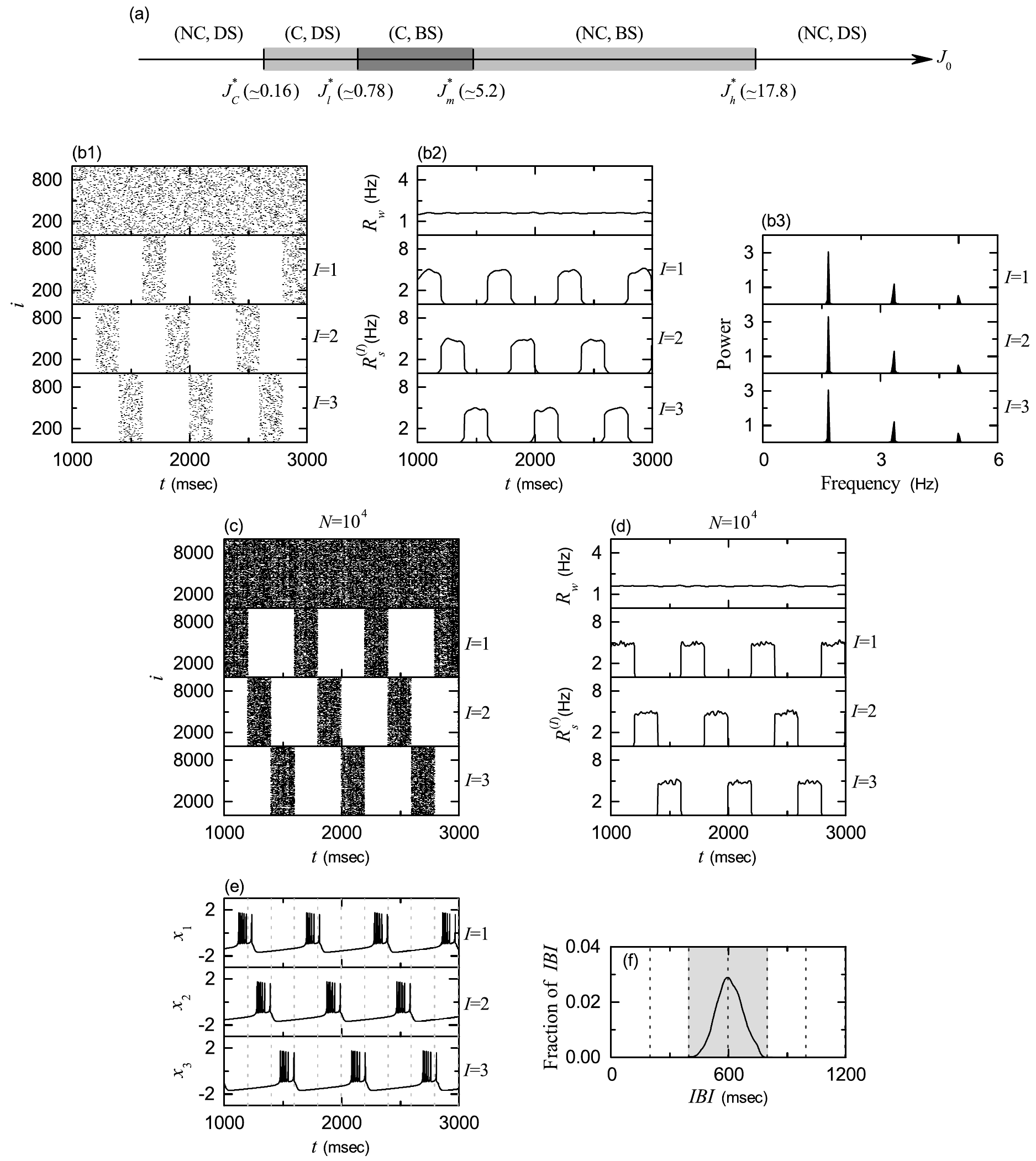}
\caption{
Emergence of 3-cluster state for $J_0=0.19$. (a) Bar diagram for the population states. C, NC, BS, and DS denote clustering, non-clustering, burst synchronization, and desynchronization, respectively. (b1) Raster plots of burst onset times in the whole population and in the $I$th cluster ($I$=1, 2, and 3). (b2) IWPBR $R_w(t)$ of the whole population and ISPBR $R_s^{(I)}(t)$ of the $I$th cluster ($I$=1, 2, and 3). (b3) One-sided power spectra of $\Delta R_s^{(I)}(t)$
[=$R_s^{(I)}(t)- \overline{R_s^{(I)}(t)}$] ($I$=1, 2, and 3) with the mean-squared amplitude normalization; the overbar denotes time average. $N=10^4$: (c) raster plots of burst onset times in the whole population and in the $I$th cluster ($I$=1, 2, and 3) and (d) IWPBR $R_w(t)$ of the whole population and ISPBR $R_s^{(I)}(t)$ of the $I$th cluster ($I$=1, 2, and 3).
(e) Time series of membrane potential $x_i(t)$ of a representative neuron $i$ in each cluster; $i$=1, 2, and 3 for $I$= 1, 2, and 3 clusters, respectively. (f) IBI histogram. Localization of IBIs in the gray region of $2~T_c < IBI < 4~ T_c$.  Vertical dotted lines in (c) and (d) denote integer multiples of the cluster period $T_c$.
}
\label{fig:C}
\end{figure}

\section{Coupling-Induced Cluster Burst Synchronization of Inhibitory HR Bursting Neurons}
\label{sec:PS}
In this section, we consider a directed Barab\'{a}si-Albert SFN, composed of $N$ inhibitory HR bursting neurons; in most cases, $N=10^3$ except for the cases of the raster plot for $N=10^4$ and the bursting order parameters $\langle {\cal{O}}_b \rangle_r$.  The synaptic coupling strengths $\{ J_{ij} \}$ are chosen from the Gaussian distribution with the mean $J_0$ and the standard deviation $\sigma_0~(=0.1)$. We investigate coupling-induced cluster burst synchronization by varying $J_0$ in the absence of noise ($D=0$).

\subsection{Emergence of Dynamical Clusterings}
\label{subsec:DC}
Figure \ref{fig:C}(a) shows a bar diagram for diverse population states. Here, C, NC, BS, and DS represent cluster, non-cluster, burst synchronization, and desynchronization, respectively.
For sufficiently small $J_0,$ non-cluster desynchronized states exist. However, when passing a critical point $J^*_c~(\simeq
0.16),$ 3-cluster states appear. As an example, we consider the case of $J_0=0.19$.
Emergence of dynamical clusterings may be well seen in the raster plot of bursting onset times which corresponds to a collection of all trains of burst onset times of individual bursting neurons. As clearly shown in Fig.~\ref{fig:C}(b), the whole population is segregated into 3 sub-populations (also called clusters); $N_I$ [number of neurons in the $I$th ($I=1,$ 2, and 3) cluster] $\simeq {\frac {N} {3}}$. Clustered busting bands appear in a successive cyclic way (i.e., $I \rightarrow I+1 \rightarrow I+2  \rightarrow I$) with the cluster period (i.e., average time interval between appearance of successive clusters) $T_c~(\simeq 199.17$ msec). Hence, in each cluster, bursting bands appear successively with the period $P~[=3~T_c~(\simeq 597.5$ msec)].

As macroscopic quantities showing the whole- and the sub-population behaviors, we employ the instantaneous whole population burst rate (IWPBR) $R_w(t)$ and the instantaneous sub-population burst rate (ISPBR) $R_s^{(I)}(t)$ ($I$=1, 2, 3) which may be
obtained from the raster plots in the whole population and in the clusters, respectively \cite{Kim1,Kim2,NN-SFN,SBS}.
To obtain a smooth IWPBR $R_w(t)$, we employ the kernel density estimation (kernel smoother) \cite{Kernel}. Each burst onset time in the raster plot is convoluted (or blurred) with a kernel function $K_h(t)$ to obtain a smooth estimate of IWPBR $R_w(t)$:
\begin{equation}
R_w(t) = \frac{1}{N} \sum_{i=1}^{N} \sum_{b=1}^{n_i} K_h (t-t_{b}^{(i)}),
\label{eq:IPBR}
\end{equation}
where $t_{b}^{(i)}$ is the $b$th burst onset time of the $i$th neuron, $n_i$ is the total number of burst onset times for the $i$th neuron, and we use a Gaussian kernel function of band width $h$:
\begin{equation}
K_h (t) = \frac{1}{\sqrt{2\pi}h} e^{-t^2 / 2h^2}, ~~~~ -\infty < t < \infty.
\label{eq:Gaussian}
\end{equation}
Throughout the paper, the band width $h$ of $K_h(t)$ is 20 msec.
The IWPBR $R_w(t)$ is shown in the top panel of Fig.~\ref{fig:C}(b2). We note that $R_w(t)$
is nearly stationary, because burst onset times in the raster plot in the whole population are nearly completely
scattered. Hence, a 3-cluster desynchronized state appears for $J_0=0.19$.

As in the case of $R_w(t)$, we get the ISPBR kernel estimate $R_s^{(I)}(t)$ by employing the Gaussian kernel function of Eq.~(\ref{eq:Gaussian}):
\begin{equation}
R_s^{(I)}(t) = \frac{1}{N_I} \sum_{i=1}^{N_I} \sum_{b=1}^{n_i^{(I)}} K_h (t-t_{b}^{(I,i)}),
\label{eq:ISPBR}
\end{equation}
where $t_{b}^{(I,i)}$ is the $b$th burst onset time of the $i$th neuron in the $I$th cluster, $n_i^{(I)}$ is the total number of burst onset times for the $i$th neuron in the $I$th cluster, and $N_I$ is the number of neurons in the $I$th cluster.
The ISPBRs $R_s^{(I)}$ of the $I$th clusters are shown in the $I=$1, 2, and 3 panels of Fig.~\ref{fig:C}(b2),  respectively.
We note that $R_s^{(I)}(t)$ shows a square-wave-like behavior. For each cluster, burst onset times in each bursting band are nearly completely scattered (i.e., nearly desynchronized), and hence a square-wave-like oscillation occurs in each $R_s^{(I)}(t)$. During the ``silent'' part (without burstings) for about $2P/3$, $R_s^{(I)}(t)=0$ (which corresponds to the bottom part), while in the bursting band for about $P/3$, $R_s^{(I)}(t)$ rapidly increases to the nearly flat top, and then decreases rapidly; $P$ corresponds to the average period of the square-wave oscillation. Through repetition of this process $R_s^{(I)}(t)$ exhibits a square-wave-like oscillation.
The sub-population bursting frequency $f_b^{(I)}$ of the ISPBR $R_s^{(I)}(t)$ $(I$=1, 2, and 3) may be obtained from the one-sided power spectra of $\Delta R_s^{(I)}(t)$ $[= R_s^{(I)}(t) - \overline{R_s^{(I)}(t)}]$ with the mean-squared amplitude normalization. The overbar represents time average and the number of data for each power spectrum is $2^{13}$.
Figure \ref{fig:C}(b3) shows power spectra of $\Delta R_s^{(I)}(t)$ ($I=$1, 2, and 3).
In the case of each sub-population (cluster), the power spectrum has a main peak at $f_b^{(I)}~(\simeq 1.67$ Hz) and its harmonics. Hence, $R_s^{(I)}(t)$ oscillates with the slow sub-population bursting frequency $f_b^{(I)}$, the reciprocal of
which corresponds to the average period $P$ of the square-wave oscillation (also corresponding to the average period for appearance of successive bursting bands in each cluster).

To examine the square-wave-like behavior more clearly, the number of HR neurons is increased from $N=10^3$ to $10^4$. In this case, raster plots in the whole population and the clusters and their corresponding IWPBR $R_w(t)$ and ISPBR  $R_s^{(I)}(t)$ are shown in Figs.~\ref{fig:C}(c) and \ref{fig:C}(d), respectively. For the whole population, burst onset times are more completely scattered, and hence the corresponding IWPBR $R_w(t)$ is more stationary. Furthermore, for each cluster, bursting bands in the raster plot show clearly the clustering structure, and the corresponding ISPBR $R_s^{(I)}(t)$ shows square-wave oscillations more clearly. Thus, for each cluster, burst onset times in bursting bands are completely scattered, and they show a desynchronized state. In this way, 3-cluster desynchronization appears for $J_0=0.19$.

We also investigate individual bursting behaviors of HR neurons in each cluster.
Figure \ref{fig:C}(e) shows a time series of a membrane potential $x_i(t)$ of a representative
neuron in each $I$th cluster ($i=$1, 2, and 3 for $I=1$, 2, and 3, respectively); the vertical dotted lines represent integer
multiples of the cluster period $T_c$.
The 1st HR neuron in the $I=1$ cluster makes burstings in the 1st clustering cycle (after the transient
time $t=10^3$ msec). We note that the duration of silent phase of the 1st neuron is about twice as long as the length of its active bursting phase. During this silent phase, the 2nd and the 3rd HR neurons in the $I=2$ and 3 clusters exhibit burstings alternately in the 2nd and the 3rd clustering cycle, respectively. In this way, individual HR neurons in each cluster
show burstings every 3rd clustering cycle. This kind of individual bursting behaviors are well shown in the IBI histograms [see Fig.~\ref{fig:C}(f)], where vertical dotted lines denote integer multiples of $T_c$.
The IBI histogram is composed of $5 \times 10^4$ IBIs, and the bin size for the histogram is 2.5 msec.
A single peak appear at $T_{peak}~(=3~T_c)$; $T_{peak}$ also corresponds to the average period $P$ for the appearance of bursting bands in each cluster. We note that IBIs are localized in a range of $2~T_c < IBI < 4~T_c$.
Based on the IBI histogram, we suggest a criterion for emergence of 3-cluster states.
Localization of IBIs in the range of $2~T_c < IBI < 4~T_c$ results in emergence of 3 clusters.

\begin{figure}
\includegraphics[width=\columnwidth]{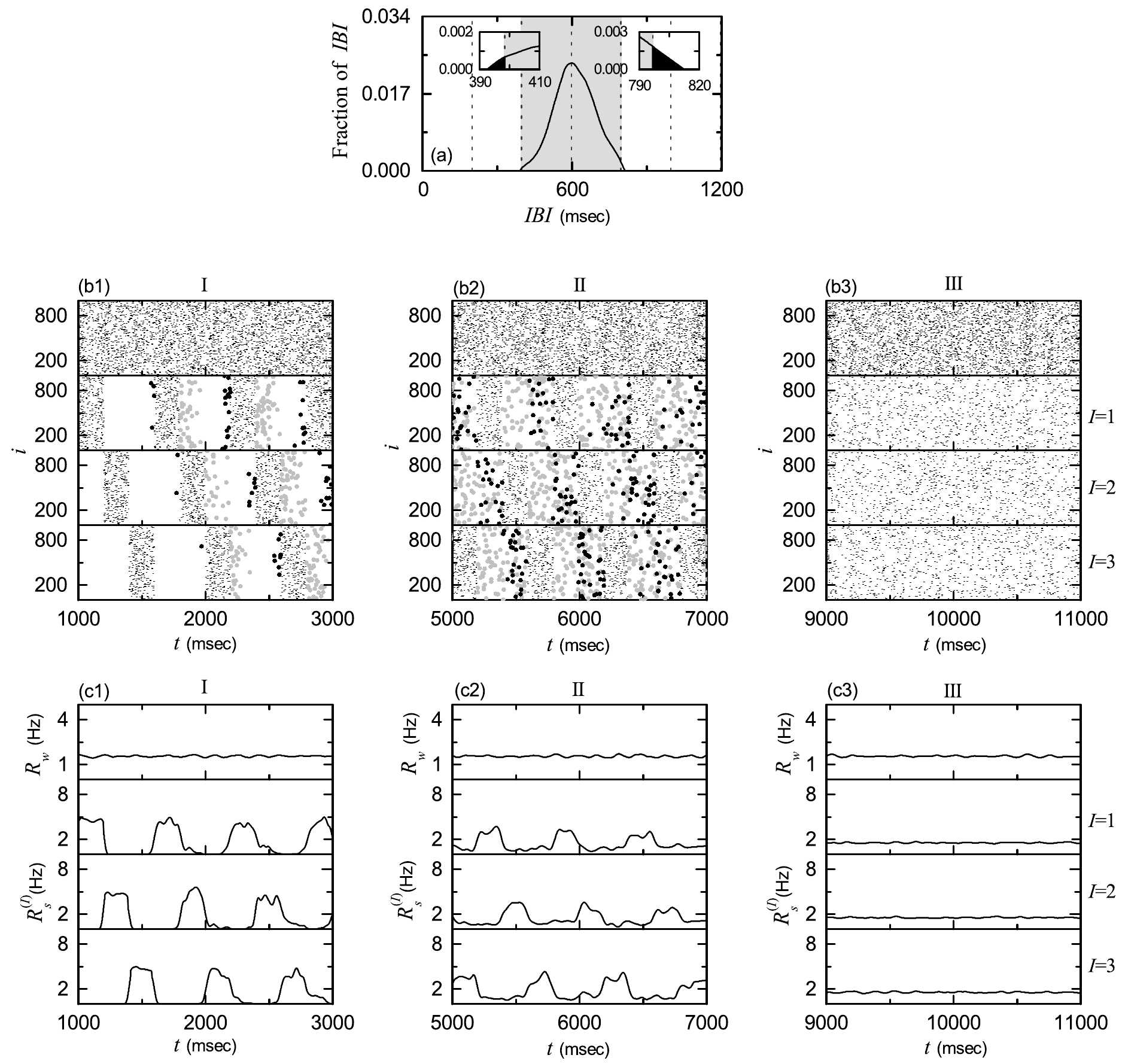}
\caption{
Break-up of 3-cluster state via intercluster hoppings for $J_0=0.13$.
(a) IBI histogram. Delocalization of IBIs through crossing the left and the right vertical boundary lines of the gray region (corresponding to $2~T_c$ and $4~T_c,$ respectively). Sequential long-term raster plots of burst onset times in the whole population and in the $I$th cluster ($I=$1, 2, and 3) in (b1) the early, (b2) the intermediate, and (b3) the final stages; in (b1) and (b2), late burstings with IBIs larger than $4~T_c$ and early burstings with IBI smaller than $2~T_c$ are plotted with gray and black dots of a little larger point size. IWPBR $R_w(t)$ of the whole population and ISPBR $R_s^{(I)}(t)$ of the $I$th cluster ($I=$1, 2, and 3) in the (c1) early, (c2) the intermediate, and (c3) the final stages.
}
\label{fig:NC1}
\end{figure}

For $J_0 < J^*_c,$ delocalization of IBIs occurs via crossing the left and/or the right boundaries (corresponding to
$2~T_c$ and $4~T_c,$ respectively). As an example, we consider the case of $J_0=0.13$. Figure \ref{fig:NC1}(a) shows
a delocalized IBI histogram. In this case, some fraction of IBIs cross both the left and the right boundaries [see the
black parts in the insets of Fig.~\ref{fig:NC1}(a)]. The fraction of IBIs above $4~T_c$ is 0.0753, while the fraction of IBIs below $2~T_c$ is 0.0146. Hence, ``late'' burstings with IBIs larger than $4~T_c$ are much more probable than ``early'' burstings with IBIs smaller than $2~T_c$. As a result of occurrence of these late and early burstings, interburst hoppings between the 3 clusters occur, which leads to break up of dynamical clusterings. Such intercluster hoppings may be well seen in sequential long-term raster plots of burst onset times in the whole population and in the $I$th ($I=1$, 2, and 3) clusters. Figures \ref{fig:NC1}(b1)-\ref{fig:NC1}(b3) show such raster plots in the early, the intermediate, and the final stages, respectively. In Figs.~\ref{fig:NC1}(b1) and \ref{fig:NC1}(b2), late and early burstings are plotted with gray and black dots of a little larger point size (=1.5), in contrast to regular burstings which are plotted with black dots of a smaller point size (=0.5).

For the initial stage in Fig.~\ref{fig:NC1}(b1), individual HR neurons in the $I$th cluster make intermittent intercluster hoppings to the nearest neighboring $(I+1)$th [$(I-1)$th] cluster due to occurrence of late (early) burstings. Thus,
bursting bands in the $I$th cluster become smeared into the nearest neighboring bursting bands belonging to the $(I+1)$th and the $(I-1)$th clusters. In this way, intermittent ``forward'' hoppings from the $I$th to the $(I+1)$th cluster and ``backward'' hoppings from the $I$th to the $(I-1)$th cluster occur through occurrence of late and early burstings, respectively.
The smearing degree of late burstings (larger gray dots) into the $(I+1)$th cluster is larger than that of early burstings
(larger black dots) into the $(I-1)$th cluster.

For the intermediate stage in Fig.~\ref{fig:NC1}(b2), one more step occurs for the intercluster hoppings
due to occurrence of 2nd late and early burstings. Hence, intercluster hoppings occur from the $I$th cluster to the $(I+1)$th and the $(I-1)$th clusters (due to the 1st late and early burstings) and then to the $(I+2)$th and the $(I-2)$th clusters (due to the 2nd late and early burstings). Thus, bursting bands in the $I$th cluster become smeared into the nearest neighboring bursting bands belonging to the $(I+1)$th and the $(I-1)$th clusters and then into the next-nearest neighboring bursting bands belonging to the $(I+2)$th and the $(I-2)$th clusters. In this way, successive 2nd forward and backward intercluster hoppings occur due to occurrence of 2nd late and early burstings, respectively. We also note that the 1st (2nd) late burstings and the 2nd (1st) early burstings are intermixed. In this intermediate stage, smeared parts into neighboring clusters are still
sparse (i.e. their densities are low in comparison with those of regular bursting bands).

As the time $t$ is further increased, 3rd late and early burstings may also occur, and then another forward (backward) intercluster hoppings from the $(I+2)$th [$(I-2)$th] to the $I$th clusters occur (i.e., return to the original $I$th cluster occurs due to the 3rd late and early burstings).
In this way, forward and backward intercluster hoppings occur in a cyclic way [$I$ $\rightarrow$ $I+1$ ($I-1$) $\rightarrow$ $I+2$ ($I-2$) $\rightarrow$ $I$] due to occurrence of successive late and early burstings. In the final stage after a sufficiently long time, intercluster hoppings between clusters are more and more intensified, which leads to complete break-up of clusters. As a result, burst onset times in the raster plots are completely scattered in a nearly uniform way, independently of $I=1,$ 2, and 3, as shown in Fig.~\ref{fig:NC1}(b3).

Figures \ref{fig:NC1}(c1)-\ref{fig:NC1}(c3) show the IWPBR $R_w(t)$ and the ISPBR $R_s^{(I)}(t)$ ($I$=1, 2, and 3), corresponding to the above raster plots in Figs.~\ref{fig:NC1}(b1)-\ref{fig:NC1}(b3).
In the initial stage in Fig.~\ref{fig:NC1}(c1), amplitudes (corresponding to heights of squares) of square-wave oscillations
are decreased and top parts of squares become less flat (i.e. they begin to wiggle).
Additionally, small-amplitude oscillations (associated with low-density smeared parts in the raster plots occurring due to forward and backward intercluster hoppings) also appear in connection with decreased square-wave oscillations

In the intermediate stage in Fig.~\ref{fig:NC1}(c2), one more step for forward and backward intercluster hoppings occurs due to 2nd late and early burstings, and hence forward and backward smearing of late and early burstings extends to the next-nearest neighboring clusters. In this case, amplitudes of square-wave oscillations are more decreased and top parts of squares also become much less flat. In this way, square-wave oscillations become broken up more and more. Additionally, small-amplitude oscillations (related to extended smeared parts in the raster plots occurring due to successive forward and backward intercluster hoppings) appear and they cover the whole range between square-wave oscillations. In this case, the amplitudes of extended small oscillations are larger than those in the initial stage.

As the time $t$ is further increased, these tendencies (e.g., decreasing tendency in amplitudes of square-wave oscillations, increasing tendency in break-up of square-wave structure, and increasing tendency in amplitudes of extended small oscillations) become intensified due to intensive forward and backward intercluster hoppings.
As a result of complete break-up of clusters, all the ISPBR $R_s^{(I)}(t)$ become nearly the same as the IWPBR $R_w(t)$, independently of $I$, and also both $R_s^{(I)}(t)$ and $R_w(t)$ are nearly stationary, as shown in Fig.~\ref{fig:NC1}(c3) for the final stage.

\begin{figure}
\includegraphics[width=\columnwidth]{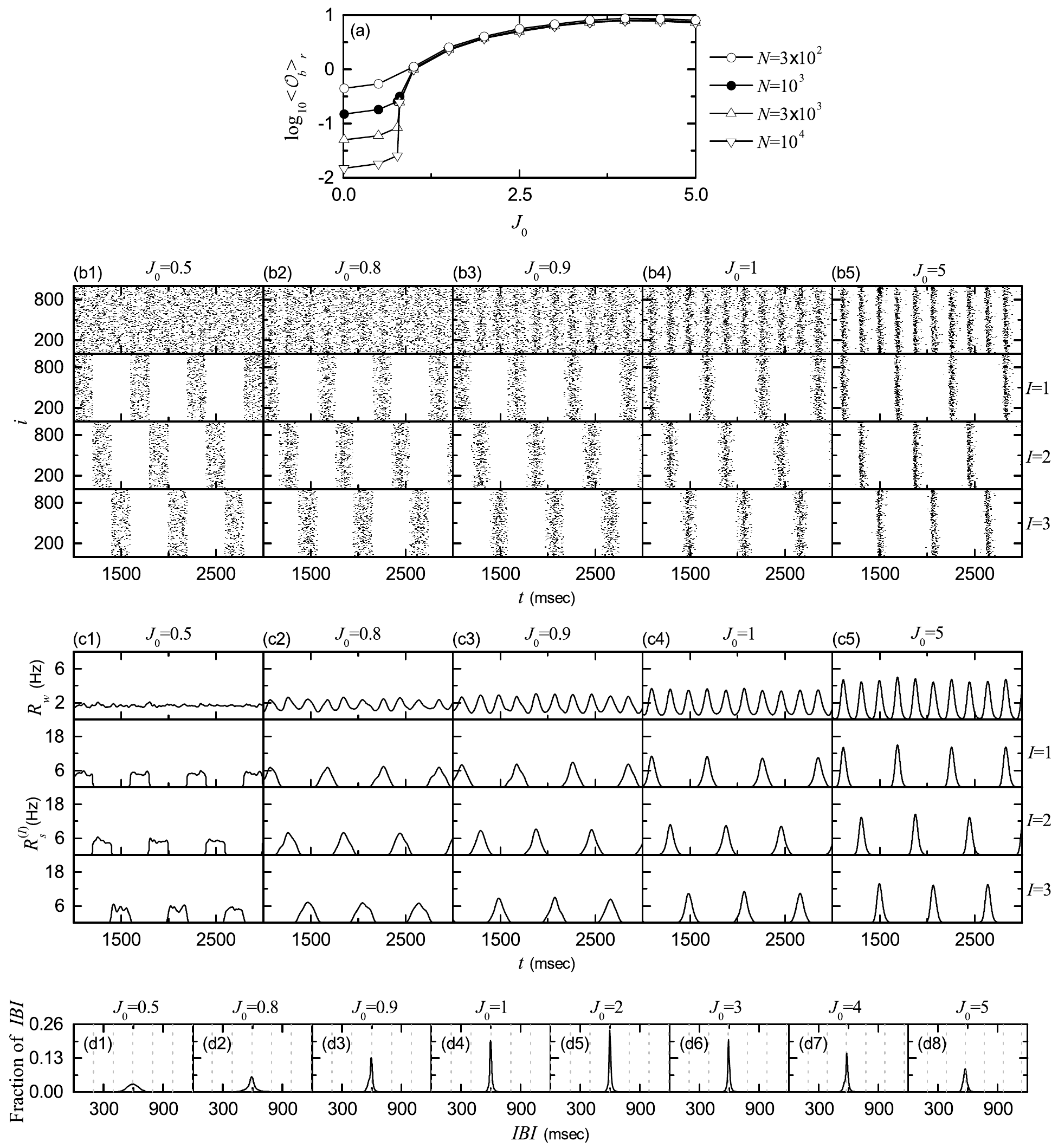}
\caption{
Emergence of 3-cluster burst synchronization. (a) Plots of thermodynamic bursting order parameter $\langle {\cal{O}}_b \rangle_r$ versus the average coupling strength $J_0$. 3-cluster desynchronization for $J_0=0.5$: (b1) raster plots of burst onset times in the whole population and in the $I$th cluster ($I$=1, 2, and 3), and (c1) IWPBR $R_w(t)$ of the whole population and ISPBR $R_s^{(I)}(t)$ of the $I$th cluster ($I$=1, 2, and 3). 3-cluster burst synchronization for various values of $J_0$:
(b2)-(b5) raster plots of burst onset times in the whole population and in the $I$th cluster ($I$=1, 2, and 3), and (c2)-(c5)
IWPBR $R_w(t)$ of the whole population and ISPBR $R_s^{(I)}(t)$ of the $I$th cluster ($I$=1, 2, and 3).
(d1)-(d8) IBI histograms for various values of $J_0$. In (d1), vertical dotted lines denote integer multiples of the cluster period $T_c$, and in (d2)-(d8) vertical dotted lines represent integer multiples of the global period $T_G$ of $R_w(t)$ ($T_G=T_c$).
}
\label{fig:CBS}
\end{figure}

\subsection{Emergence of Cluster Burst Synchronization}
\label{subsec:CBS}
As the average coupling strength $J_0$ is increased and passes a threshold, a transition from cluster desynchronization to cluster burst synchronization occurs. In a desynchronized case,  burst onset times are completely scattered in the raster plot in the whole population [e.g., see the top panel of Fig.~\ref{fig:CBS}(b1)]. On the other hand, in the case of burst synchronization, bursting stripes (composed of burst onset times and representing burst synchronization) appear successively in the raster plot in the whole population [e.g., see the top panels of Figs.~\ref{fig:CBS}(b2)-\ref{fig:CBS}(b5)].

Recently, we introduced a realistic bursting order parameter, based on $R_w(t)$, for describing transition from desynchronization to burst synchronization \cite{Kim2}.
The mean square deviation of $R_w(t)$,
\begin{equation}
{\cal{O}}_b \equiv \overline{(R_w(t) - \overline{R_w(t)})^2},
 \label{eq:Order}
\end{equation}
plays the role of an order parameter ${\cal{O}}_b$; the overbar represents time average. This bursting order parameter may be regarded as a thermodynamic measure because it concerns just the macroscopic IWPBR $R_w(t)$ without any consideration between $R_w(t)$ and microscopic individual burst onset times. As $N$ (number of HR neurons in the whole population) is increased, $R_w(t)$ exhibits more regular oscillations in the case of burst synchronization, while it becomes more stationary in the case of desynchronization. Hence, in the thermodynamic limit of $N \rightarrow \infty$, the bursting order parameter ${\cal{O}}_b$, representing time-averaged fluctuations of $R_w(t)$ from its time-averaged mean, approaches a non-zero (zero) limit value for the synchronized (desynchronized) state. In this way, the bursting order parameter ${\cal{O}}_b$ can determine whether population states are synchronized or desynchronized.

Figure \ref{fig:CBS}(a) shows a plot of $\log_{10} \langle {\cal{O}}_b \rangle_r$ versus $J_0$.
In each realization, we discard the first time steps of a trajectory as transients for $10^3$ msec, and then we numerically compute ${\cal{O}}_b$ by following the trajectory for $3 \times 10^4$ msec. Hereafter, $\langle \cdots \rangle_r$ denotes an average over 20 realizations.
For $J_0 < J_l^* (\simeq 0.78)$, the bursting order parameter $\langle {\cal{O}}_b \rangle_r$ tends to zero with increasing $N$.
On the other hand, when passing $J_l^*$ a transition to burst synchronization occurs, because $\langle {\cal{O}}_b \rangle_r$
approaches a non-zero limit value. Consequently, for $J_0 > J_l^*$ burst synchronization occurs in the whole population due to a constructive role of synaptic inhibition to favor the burst synchronization.

We consider specific examples of cluster desynchronization and cluster burst synchronization.
Figures \ref{fig:CBS}(b1) and \ref{fig:CBS}(c1) show an example of cluster desynchronization for $J_0=0.5$, as in the case of
$J_0=0.13$ in Figs.~\ref{fig:C}(b1) and \ref{fig:C}(b2). For this cluster desynchronized state, burst onset times are completely scattered in bursting bands in each cluster, the corresponding ISPBR $R_s^{(I)}(t)$ exhibit square-wave oscillations, and the IWPBR $R_w(t)~[\simeq {\frac {1} {3}} \sum_{I=1}^3 R_s^{(I)}(t)$] in the whole population becomes nearly stationary.
Four examples for cluster burst synchronization are given for $J_0=0.8,$ 0.9, 1.0 and 5.0.
In the case of $J_0=0.8$, bursting stripes begin to appear successively in the raster plot of burst onset times in the whole population [see the top panel of Fig.~\ref{fig:CBS}(b2)], and the corresponding IWPBR $R_w(t)$ also begins to exhibit small-amplitude regular oscillations, as shown in the top panel of Fig.~\ref{fig:CBS}(c2). The whole population is segregated into 3 clusters. Bursting stripes in each cluster appear successively every 3rd global cycle of $R_w(t)$, as shown in the $I=$ 1, 2, and 3 panels of Figs.~\ref{fig:CBS}(b2). The ISPBRs $R_s^{(I)}$ of the $I$th clusters are shown in the $I=$1, 2, and 3 panels of Figs.~\ref{fig:CBS}(c2), respectively. They exhibit regular oscillations with the sub-population bursting frequency $f_b^{(I)}~(\simeq 1.67$ Hz) which corresponds to $\frac {f^{(w)}_b} {3}$ [$f^{(w)}_b$: whole-population bursting frequency of $R_w(t)$]. With increasing $J_0,$ cluster burst synchronization gets better, as shown in the cases of $J_0=0.9,$ 1.0, and 5.0.
Bursting stripes in the raster plots (in the whole population and the clusters) become clearer (i.e. less smeared) [see Figs.~\ref{fig:CBS}(b3)-\ref{fig:CBS}(b5)] and the amplitudes of $R_w(t)$ and $R_s^{(I)}(t)$ become larger
[see Figs.~\ref{fig:CBS}(c3)-\ref{fig:CBS}(c5)].

We also investigate individual bursting behaviors of HR neurons in terms of IBIs.
Figures \ref{fig:CBS}(d1)-\ref{fig:CBS}(d8) show IBI histograms for various values of $J_0$.
Each IBI histogram is composed of $5 \times 10^4$ IBIs and the bin size for the histogram is 2.5 msec.
Vertical dotted lines in the IBI histograms represent integer multiples of the cluster period $T_c$; in the case of cluster burst synchronization, the value of $T_c$ is equal to that of the global period $T_G$ of $R_w(t)$.
In all cases where 3-cluster states exist, single peaks appear at $3~T_c$, and IBIs are localized in a range of $2~T_c < IBI < 4~T_c$, as in the case of $J_0=0.19$ in Fig.~\ref{fig:C}(f).
In the desynchronized case of $J_0=0.5,$ its IBI histogram is broad due to incoherent synaptic inputs.
When passing the lower threshold $J^*_i~(\simeq 0.78)$, a transition to burst synchronization occurs, and then IBI histograms
begin to be sharp due to coherent synaptic inputs. As $J_0$ is further increased, the peaks of the IBI histograms become sharper
due to increase in coherent synaptic inputs. A maximum height of the peak appears near $J_0=2.0$, and then it begins to decrease. Thus, the peak for $J_0=5.0$ becomes broader, because $J_0=5.0$ is close to an intermediate threshold $J^*_m(\simeq 5.2$) where break-up of 3 clusters occurs (this point is explained in details in the following subsection).

\begin{figure}
\includegraphics[width=\columnwidth]{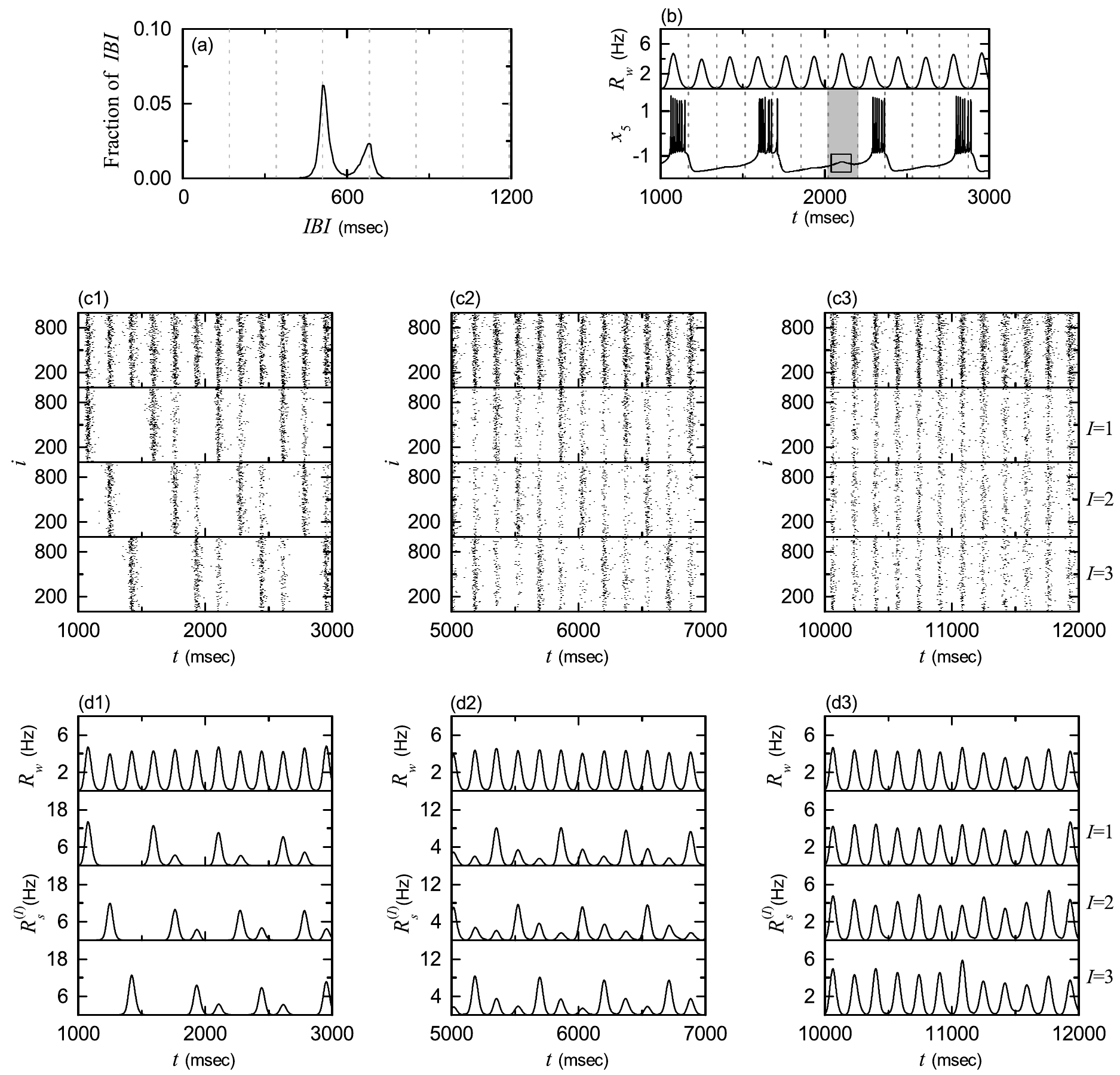}
\caption{Break-up of 3 clusters via intercluster hoppings for $J_0=10$.
(a) Double-peaked IBI histogram. Vertical dotted lines denote integer multiples of the global period $T_G~(\simeq 170.5$ msec) of $R_w(t)$. (b) Time series of membrane potential $x_5(t)$ for the 5th HR neuron in the 1st ($I=1$) cluster. For reference, the IWPBR $R_w(t)$ of the whole population is shown in the top panel. The 5th HR neuron exhibits a bursting at the 4th cycle of $R_w(t)$ rather than at its 3rd cycle where it shows a hopping instead of bursting, as shown inside the small box in the gray region (corresponding to the 3rd cycle). Sequential long-term raster plots of burst onset times in the whole population and
in the $I$th cluster ($I=$1, 2, and 3) in (c1) the early, (c2) the intermediate, and (c3) the final stages. IWPBR
$R_w(t)$ of the whole population and ISPBR $R_s^{(I)}(t)$ of the $I$th cluster ($I=$1, 2, and 3) in the (d1) early, (d2) the intermediate, and (d3) the final stages.
}
\label{fig:ICH}
\end{figure}

\subsection{Break-up of Cluster Burst Synchronization via Intercluster Hoppings}
\label{subsec:ICH}
As $J_0$ is further increased and passes an intermediate threshold $J_m^*$ $(\simeq 5.2$),
3-cluster burst synchronization breaks up into (non-cluster) burst synchronization without dynamical clusterings through intercluster hoppings. As an example, we consider the case of $J_0=10$.

Figure \ref{fig:ICH}(a) shows the IBI histogram with two peaks at $3~T_G$ and $4~T_G$ [$T_G~(\simeq 170.5$ msec): global period of $R_w(t)$]. For $J_0 < J^*_m,$ only single peak appears at $3~T_G$ [i.e., individual HR neurons exhibit burstings every 3rd global cycle of $R_w(t)$], as shown in Figs.~\ref{fig:CBS}(d2) - \ref{fig:CBS}(d8). As $J_0$ approaches the threshold $J^*_m$, this peak becomes broad along with decrease in its height. After passing $J^*_m$, individual HR neurons begin to exhibit burstings intermittently at a 4th cycle of $R_w(t)$ through burst skipping at its 3rd cycle. Here, 3rd and 4th cycles of $R_w(t)$ refer to ones counted just after the latest burstings (e.g., see the example given below).
An example for the 5th neuron in the 1st ($I=1$) cluster is given in Fig.~\ref{fig:ICH}(b). A burst skipping occurs in the small box in the gray region [corresponding to a 3rd cycle of $R_w(t)$], and then another bursting appears at its 4th cycle; for reference, $R_w(t)$ is shown on the top panel and vertical dotted lines represent global cycles of $R_w(t)$. Thus, in addition to the major peak at $3~T_G$, a new minor peak appears at $4~T_G$ in Fig.~\ref{fig:ICH}(a).
Then, some fraction of IBIs with larger than $4~T_G$ appear (i.e., late burstings occur) in contrast to the case of cluster burst synchronization where IBIs are localized in a range of $2~T_G < IBI < 4~T_G$. In this case, delocalization of IBIs occurs by crossing just the right boundary (corresponding to $4~T_c$), which is in contrast to the case of $J_0=0.13$ where both the left and the right boundaries are crossed.

Due to appearance of delocalized IBIs larger than $4~T_G$ (i.e., because of occurrence of late burstings), only forward intercluster hoppings occur, in contrast to the case of $J_0=0.13$ where both forward and backward intercluster hoppings take place due to occurrence of late and early burstings, respectively [see Figs.~\ref{fig:NC1}(b1)-\ref{fig:NC1}(b3)].
Forward intercluster hoppings between the 3 clusters may be well seen in sequential long-term raster plots of burst onset times in the whole population and in the $I$th ($I=1$, 2, and 3) clusters. Figures \ref{fig:ICH}(c1)-\ref{fig:ICH}(c3) show such raster plots, corresponding to (c1) the early, (c2) the intermediate, and (c3) the final stages. For the initial stage in Fig.~\ref{fig:ICH}(c1), individual HR neurons in the $I$th cluster make intermittent intercluster hoppings to the nearest neighboring $(I+1)$th cluster [i.e., neurons in the $I$th cluster exhibit intermittent burstings at a 4th cycle of $R_w(t)$ (along with regular burstings of neurons in the $(I+1)$th cluster) due to burst skipping at its 3rd cycle]. As a result, additional bursting stripes (composed of intermittent burstings occurring at a 4th cycle of $R_w(t)$ due to burst skipping at its 3rd cycle) appear next to the regular bursting stripes in the raster plot for the $I$th cluster. These additional bursting stripes in the $I$th cluster are vertically aligned with regular bursting stripes in the $(I+1)$th cluster. In this way, intermittent hoppings from the $I$th to the $(I+1)$th clusters occur.

For the intermediate stage in Fig.~\ref{fig:ICH}(c2), one more step occurs for the intercluster hoppings
due to a 2nd burst skipping, and hence intercluster hoppings occur from the $I$th cluster to the $(I+1)$th cluster (due to a 1st burst skipping) and then to the $(I+2)$th cluster (due to a 2nd burst skipping). Consequently, two successive additional bursting stripes (consisting of intermittent burstings occurring at a 4th  cycle of $R_w(t)$ due to the 1st and 2nd burst skippings) appear next to the regular bursting stripes in the raster plot in the $I$th cluster. These two additional bursting stripes are vertically aligned with regular bursting stripes in the $(I+1)$th and the $(I+2)$th clusters. Consequently, for each cluster, bursting stripes appear at every cycle of $R_w(t)$ in the raster plot, like the case of whole population, although additional bursting stripes (formed due to burst skippings) are still sparse (i.e. their densities are low in comparison with those of regular bursting stripes).

As the time $t$ is further increased, a 3rd burst skipping may also occur, and then another intercluster hopping from the $(I+2)$th to the $I$th clusters occurs (i.e., return to the original $I$th cluster occurs due to a 3rd burst skipping). In this way, intercluster hoppings occur in a cyclic way ($I$ $\rightarrow$ $I+1$ $\rightarrow$ $I+2$ $\rightarrow$ $I$) due to successive burst skippings. After a sufficiently long time, in the final stage in Fig.~\ref{fig:ICH}(c3), intercluster hoppings between clusters are more and more intensified, which leads to complete break-up of clusters. As a result, density of all bursting stripes becomes nearly the same, independently of $I=1,$ 2, and 3. We also note that, in spite of break-up of clusters, burst synchronization persists in the whole population, because bursting stripes appear successively in the rater plot in the whole population.

Figures \ref{fig:ICH}(d1)-\ref{fig:ICH}(d3) show the IWPBR $R_w(t)$ and the ISPBR $R_s^{(I)}(t)$, corresponding to the above raster plots in Figs.~\ref{fig:ICH}(c1)-\ref{fig:ICH}(c3). In the initial stage in Fig.~\ref{fig:ICH}(d1), smaller-amplitude oscillations [corresponding to lower-density additional bursting stripes appearing due to burst skippings at regular 3rd cycles of $R_w(t)$] appear next to the regular oscillations [occurring at every 3rd cycle of $R_w(t)$] in each $I$th ($I=1$, 2, and 3) case. In the intermediate stage in Fig.~\ref{fig:ICH}(d2), one more step for intercluster hoppings occurs due to 2nd burst skippings, and hence two successive smaller-amplitude oscillations appear next to the regular oscillations in each $I$th ($I=1$, 2, and 3) case. Then, for each $I$th cluster $R_s^{(I)}(t)$ makes oscillations at every cycle of $R_w(t)$, although its amplitudes vary depending on the cycles of $R_w(t)$. As the time $t$ is further increased, these amplitudes tend to become nearly the same due to intensified intercluster hoppings, as shown in Fig.~\ref{fig:ICH}(d3) for the final stage. Consequently, all the ISPBRs $R_s^{(I)}(t)$ become nearly the same as the IWPBR $R_w(t)$, independently of $I$, because of complete break-up of clusters.

\begin{figure}
\includegraphics[width=0.9\columnwidth]{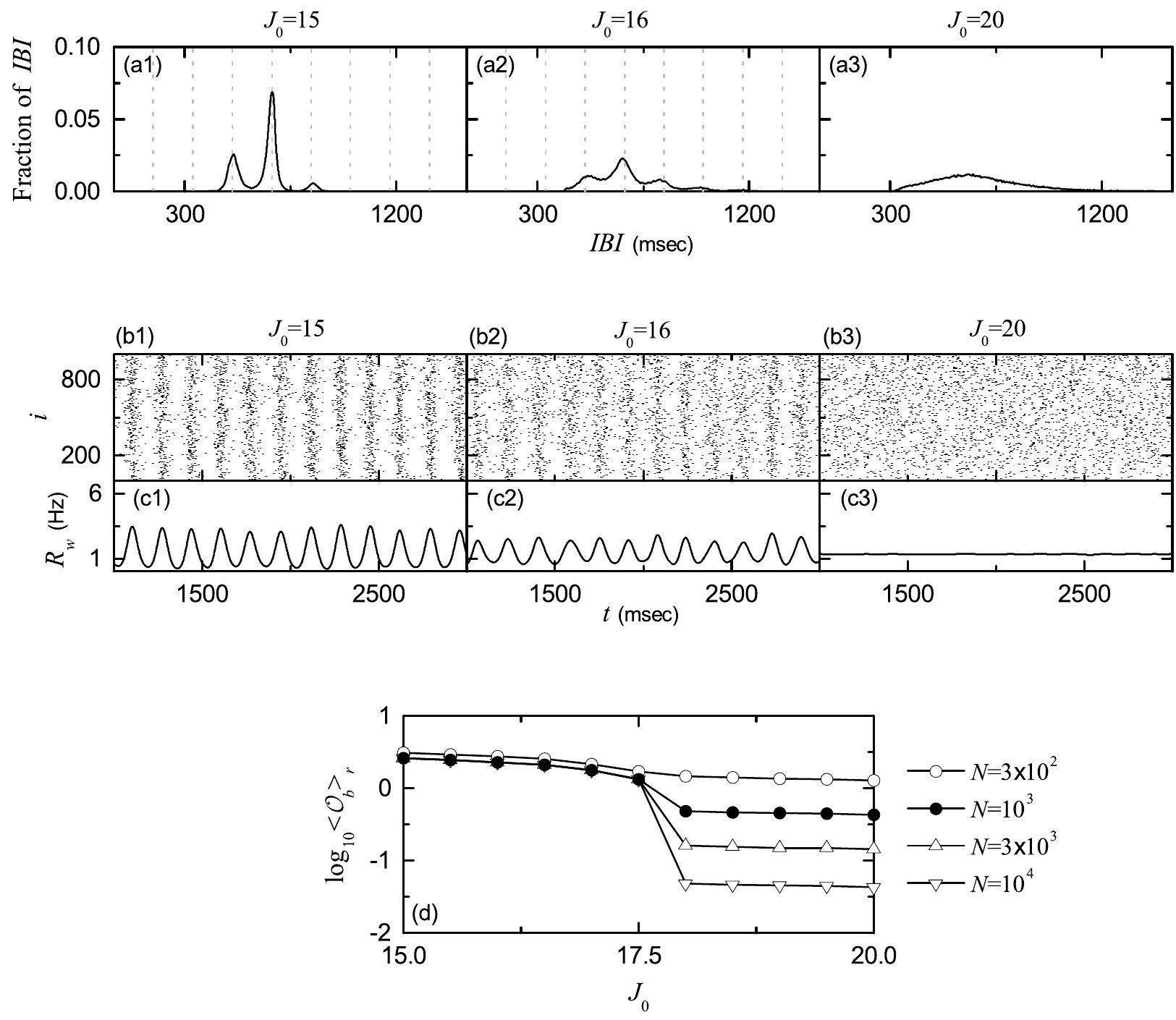}
\caption{
Transition from burst synchronization to desynchronization. IBI histograms for $J_0=$ (a1) 15, (a2) 16, and (a3) 20. Vertical dotted lines in (a1) and (a2) denote integer multiples of the global period $T_G$ of $R_w(t)$.
Raster plots of burst onset times for $J_0=$ (b1) 15, (b2) 16, and (b3) 20. IWPBR $R_w(t)$ of the whole population for $J_0=$ (c1) 15, (c2) 16, and (c3) 20. (d) Plots of thermodynamic bursting order parameter $\langle {\cal {O}}_b \rangle_r$ versus the average coupling strength $J_0$.
}
\label{fig:DS}
\end{figure}
So far, we consider the case of $J_0=10$ where (non-cluster) burst synchronization without dynamical clusterings appears via
intercluster hoppings which occur due to burst skippings. With increase in $J_0$ from 10, another type of bursting skippings begin to occur at 4th cycles of $R_w(t)$, in addition to the above skippings at 3rd cycles for $J_0=10$. Figure \ref{fig:DS}(a1) shows the IBI histogram for $J_0=15$. When compared with the IBI histogram for $J_0=10$ in Fig.~\ref{fig:ICH}(a), the height of the peak at $4~T_G$ is so much increased, and hence its height becomes higher than that of the decreased peak at $3~T_G$. As a result, the peak at $4~T_G$ becomes a major one. Furthermore, a new smaller peak appears at $5~T_G$ due to intermittent burst skippings at 4th cycles of $R_w(t)$. Thus, the IBI histogram for $J_0=15$ consists of 3 peaks at $3~T_G$, $4~T_G$, and $5~T_G$.
Figures \ref{fig:DS}(b1) and \ref{fig:DS}(c1) show the raster plot in the whole population and the corresponding IWPBR kernel estimate $R_w(t)$ for $J_0=15$. In comparison with the case of $J_0=10$ in Figs.~\ref{fig:ICH}(c3) and \ref{fig:ICH}(d3), due to a destructive role of synaptic inhibition to spoil the burst synchronization, burst stripes become more smeared, and amplitudes of $R_w(t)$ become smaller. Consequently, the degree of (non-cluster) burst synchronization becomes worse.

As $J_0$ is further increased, this kind of tendency for burst skippings is intensified. As an example, see the case of $J_0=16$. The IBI histogram is shown in Fig.~\ref{fig:DS}(a2). In comparison with the IBI histogram in Fig.~\ref{fig:DS}(a1) for $J_0=15,$ heights of both peaks at $3~T_G$ and $4~T_G$ are decreased, while the height of the
peak at $5~T_G$ is a little increased. Additionally, a new small peak appears at $6~T_G$. In this way, the IBI distribution becomes broad. When compared with the case of $J_0=15$, bursting stripes become more smeared and amplitudes of $R_w(t)$ are decreased, as shown in Figs.~\ref{fig:DS}(b2) and \ref{fig:DS}(c2), respectively. In this way, with increasing $J_0$  (non-cluster) burst synchronization becomes more and more worse.

Eventually, when passing a higher threshold $J^*_h~(\simeq 17.8)$, a transition to desynchronization occurs. Consequently, for $J_0 > J^*_h$ desynchronized states appear, as shown in the case of $J_0=20$. In this case, the IBI histogram is so broad and has just a central maximum via merging of peaks. Burst onset times in the raster plot are completely scattered without forming any bursting stripes, and the corresponding IWPBR kernel estimate $R_w(t)$ becomes nearly stationary [see Figs.~\ref{fig:DS}(b3) and \ref{fig:DS}(c3), respectively]. This type of transition from burst synchronization to desynchronization may also be well described in terms of the bursting order parameter ${\cal{O}}_b$ of Eq.~(\ref{eq:Order}). Figure \ref{fig:DS}(d) shows a plot of $\log_{10} \langle {\cal{O}}_b \rangle_r$ versus $J_0$. As $N$ is increased, the bursting order parameter $\langle {\cal{O}}_b \rangle_r$ approaches a non-zero limit value for $J_0 < J_h^* (\simeq 17.8)$, and hence (non-cluster) burst synchronization occurs. On the other hand, when passing $J_h^*$ a transition to (non-cluster) desynchronization occurs, because $\langle {\cal{O}}_b \rangle_r$ tends to zero with increasing $N$. Consequently, for $J_0 > J_h^*$ (non-cluster) desynchronized states appear due to a destructive role of synaptic inhibition to spoil the burst synchronization.

\subsection{Characterization of Burst Synchronization}
\label{subsection:CHBS}
We characterize burst synchronization in the range of $J^*_l < J_0 < J^*_h$ by employing a statistical-mechanical bursting measure $M_b$ \cite{Kim2}. In the case of burst synchronization, bursting stripes appear successively in the raster plot of burst onset times in the whole population. The bursting measure $M^{(b)}_i$ of the $i$th bursting stripe is defined by the product of the occupation degree $O^{(b)}_i$ of burst onset times (representing the density of the $i$th bursting stripe) and the pacing degree $P^{(b)}_i$ of burst onset times (denoting the degree of phase coherence between burst onset times in the $i$th bursting stripe):
\begin{equation}
M^{(b)}_i = O^{(b)}_i \cdot P^{(b)}_i.
\label{eq:BMi}
\end{equation}
The occupation degree $O^{(b)}_i$ of burst onset times in the $i$th bursting stripe is given by the fraction of bursting neurons:
\begin{equation}
   O^{(b)}_i = \frac {N_i^{(b)}} {N},
\label{eq:OD}
\end{equation}
where $N_i^{(b)}$ is the number of bursting neurons in the $i$th bursting stripe.
In the case of full burst synchronization, all bursting neurons exhibit burstings in each bursting stripe in the raster plot of burst onset times, and hence the occupation degree $O_i^{(b)}$ in each bursting stripe becomes 1. On the other hand, in the case of sparse burst synchronization, only some fraction of bursting neurons show burstings in each bursting stripe, and hence the occupation degree $O_i^{(b)}$ becomes less than 1.
In our case of burst synchronization, $O^{(b)}_i <1$ in the range of $ J^*_l < J_0 < J^*_h$, and hence sparse
burst synchronization occurs.

The pacing degree $P^{(b)}_i$ of burst onset times in the $i$th bursting stripe can be determined in a statistical-mechanical way by taking into account their contributions to the macroscopic IWPBR $R_w(t)$.
Central maxima of $R_w(t)$ between neighboring left and right minima of $R_w(t)$ coincide with centers of bursting stripes in the raster plot. A global cycle starts from a left minimum of $R_w(t)$, passes a maximum, and ends at a right minimum.
An instantaneous global phase $\Phi^{(b)}(t)$ of $R_w(t)$ was introduced via linear interpolation in the region forming a global cycle (for more details, refer to Eqs.~(14) and (15) in \cite{Kim2}).  Then, the contribution of the $k$th microscopic burst onset time in the $i$th bursting stripe occurring at the time $t_k^{(b)}$ to $R_w(t)$ is
given by $\cos \Phi^{(b)}_k$, where $\Phi^{(b)}_k$ is the global phase at the $k$th burst onset time [i.e., $\Phi^{(b)}_k \equiv \Phi^{(b)}(t_k^{(b)})$]. A microscopic burst onset time makes the most constructive (in-phase)
contribution to $R_w(t)$ when the corresponding global phase $\Phi^{(b)}_k$ is $2 \pi n$ ($n=0,1,2, \dots$), while it makes the most destructive (anti-phase) contribution to $R_w(t)$ when $\Phi^{(b)}_k$
is $2 \pi (n-1/2)$. By averaging the contributions of all microscopic burst onset times in the $i$th bursting stripe to $R_w(t)$, we obtain the pacing degree of burst onset times in the $i$th stripe:
\begin{equation}
 P^{(b)}_i ={ \frac {1} {B_i}} \sum_{k=1}^{B_i} \cos \Phi^{(b)}_k,
\label{eq:PD}
\end{equation}
where $B_i$ is the total number of microscopic burst onset times in the $i$th stripe.
By averaging $M_i^{(b)}$ of Eq.~(\ref{eq:BMi}) over a sufficiently large number $N_b$ of bursting stripes, we obtain the realistic statistical-mechanical bursting measure $M_b$, based on the IWPBR $R_w(t)$:
\begin{equation}
M_b =  {\frac {1} {N_b}} \sum_{i=1}^{N_b} M^{(b)}_i.
\label{eq:BM}
\end{equation}
We follow $3 \times 10^3$ bursting stripes in each realization and get $\langle M_b \rangle_r$ via average over 20 realizations.

\begin{figure}
\includegraphics[width=0.8\columnwidth]{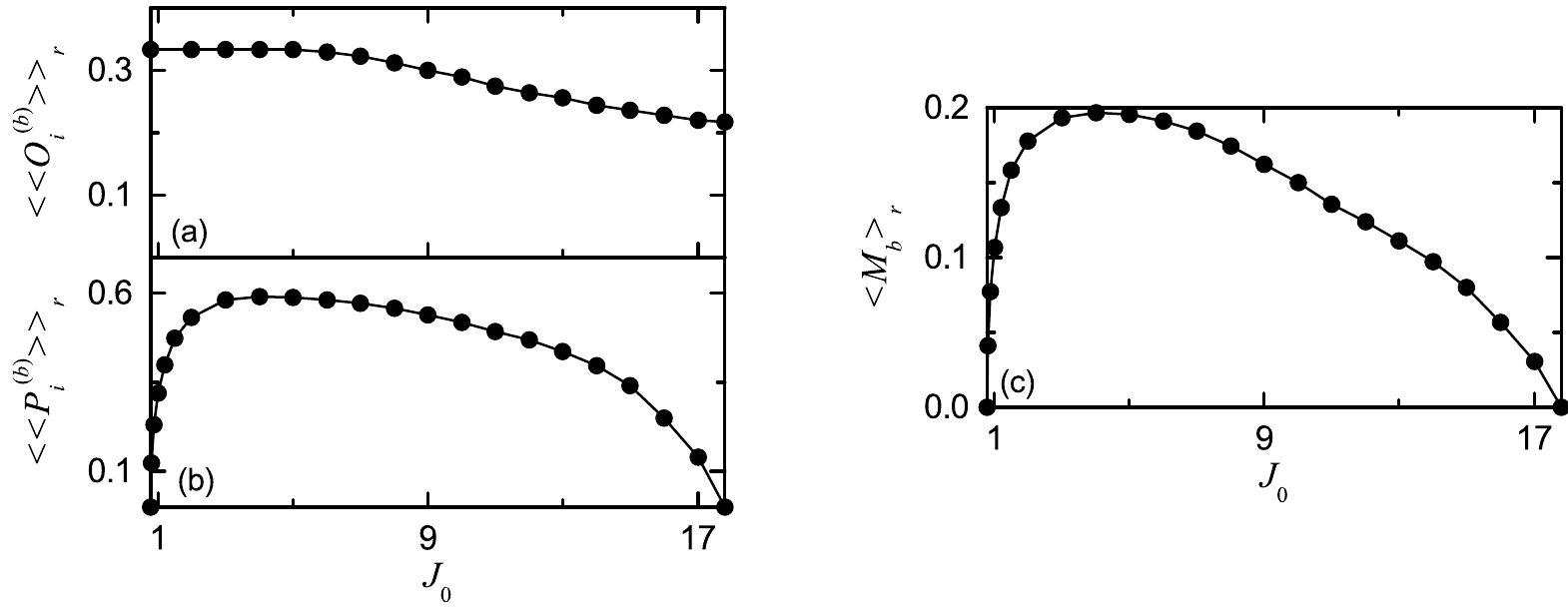}
\caption{
Characterization of burst synchronization. Plots of (a) the average occupation degree $\langle \langle O_i^{(b)} \rangle \rangle_r$, (b) the average pacing degree $\langle \langle P_i^{(b)} \rangle \rangle_r$, and (c) the statistical-mechanical bursting measure $\langle M_b \rangle_r$  versus the average coupling strength $J_0$.
}
\label{fig:Char}
\end{figure}

Figures \ref{fig:Char}(a)-\ref{fig:Char}(c) show the average occupation degree $\langle \langle O_i^{(b)} \rangle \rangle_r$, the average pacing degree $\langle \langle P_i^{(b)} \rangle \rangle_r$, and the statistical-mechanical bursting measure $\langle M_b \rangle_r$, respectively. In the case of 3-cluster burst synchronization in the range of $J^*_l~(\simeq 0.78) < J_0 < J^*_m~(\simeq 5.2)$, $\langle \langle O_i^{(b)} \rangle \rangle_r$ (denoting the density of bursting stripes in the raster plot) is $\frac {1} {3}$, because individual HR neurons exhibit burstings every 3rd cycle of $R_w(t)$. However, for $J_0 > J^*_m$, $\langle \langle O_i^{(b)} \rangle \rangle_r$ decreases slowly to a limit value $(\simeq 0.217)$, due to burst skippings [e.g., see IBI histograms Figs.~\ref{fig:DS}(a1) and \ref{fig:DS}(a2)].
The average pacing degree $\langle \langle P_i^{(b)} \rangle \rangle_r$ represents well the average degree of phase coherence in bursting stripes in the raster plot of burst onset times [e.g., see Figs.~\ref{fig:CBS}(b2)-\ref{fig:CBS}(b8) and Fig.~\ref{fig:DS}(b1) and \ref{fig:DS}(b2)].
As $J_0$ is increased from $J^*_l$, $\langle \langle P_i^{(b)} \rangle \rangle_r$ increases rapidly to a maximum $(\simeq 0.591)$ for $J_0 \simeq 4.0$ (i.e., the degree of 3-cluster burst synchronization increases rapidly after its appearance). Then, for $J_0 > 4$ it decreases to zero at the higher transition point $J^*_h~(\simeq 17.8)$ (i.e., decrease in $\langle \langle P_i^{(b)} \rangle \rangle_r$ begins a little before break-up of 3-cluster burst synchronization for $J_0=J^*_m$, and then $\langle \langle P_i^{(b)} \rangle \rangle_r$ decreases smoothly to zero, due to complete overlap of sparse bursting stripes).
Through averaging product of the occupation and the pacing degrees of burst onset times over sufficiently large number of bursting stripes in each realization, the statistical-mechanical bursting measure $\langle M_b \rangle_r$ is obtained. Since the variation in $\langle \langle O_i^{(b)} \rangle \rangle_r$ is small, $\langle M_b \rangle_r$ behaves like the case of $\langle \langle P_i^{(b)} \rangle \rangle_r$. With increasing $J_0$ from $J^*_l$, $\langle M_b \rangle_r$ increases rapidly to a maximum $(\simeq 0.197)$ for $J_0 \simeq 4.0,$ and then, for $J_0 > 4$ it decreases slowly to zero at the higher transition point $J^*_h$

\begin{figure}
\includegraphics[width=0.6\columnwidth]{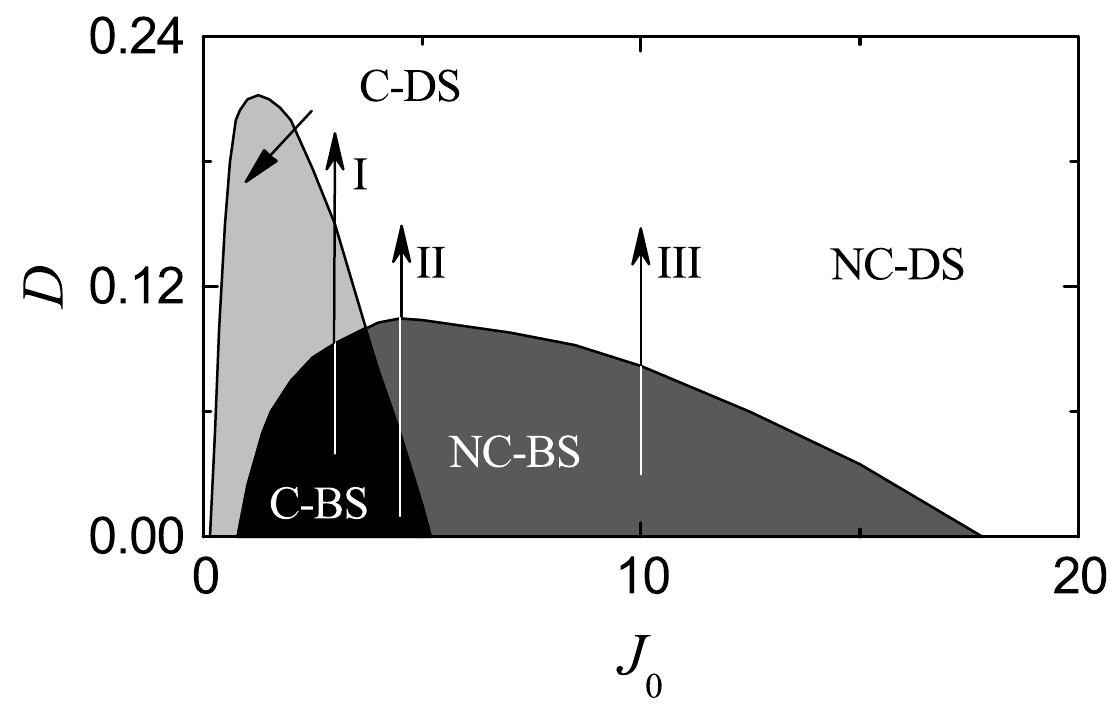}
\caption{
State diagram in the $J_0-D$ plane. Four types of population states exist. 3-cluster burst synchronization occurs in the black region, denoted by $C-BS,$ while non-cluster burst synchronization occurs in the dark gray region, represented by $NC-BS.$ Cluster desynchronization appears in the gray region, denoted by $C-DS,$ while non-cluster desynchronization appears
in the remaining white region, represented by $NC-DS.$ Vertical arrows ($I$, $II$, and $III$) represent routes for
$J_0=3$, 4.5, and 10 where effects of stochastic noise are studied.
}
\label{fig:SD}
\end{figure}

\section{Effects of Stochastic Noise on Cluster Burst Synchronization}
\label{sec:Noise}
In this section, we study the effects of stochastic noise on burst synchronization by changing the noise intensity $D$.
First, we obtain the state diagram in the $J_0-D$ plane, which is shown in Fig.~\ref{fig:SD}. Four types of population states
exist. 3-cluster states appear in the gray region, denoted by $C$, on the left side. Also, burst synchronization occurs in the dark gray region, represented by $BS,$ on the right side. In the intersection region, shaded in black and denoted by $C-BS,$ between the cluster and the burst synchronization regions, 3-cluster burst synchronization occurs.
On the other hand, in the remaining regions of the cluster and the burst synchronization regions, cluster desynchronization and non-cluster burst synchronization occurs, respectively; these remaining regions are denoted by $C-DS$ and $NC-BS,$ respectively. Outside these cluster and burst synchronization regions, non-cluster desynchronization occurs in a region denoted by $NC-DS$.

Next, we investigate the effects of noise on 3-cluster burst synchronization and intercluster hoppings (studied in the above section for $D=0$) by increasing the noise intensity $D$ along the 3 routes for $J_0=$3, 4.5, and 10, denoted by vertical arrows
($I$, $II,$ and $III$) in the state diagram of Fig.~\ref{fig:SD}.

\begin{figure}
\includegraphics[width=\columnwidth]{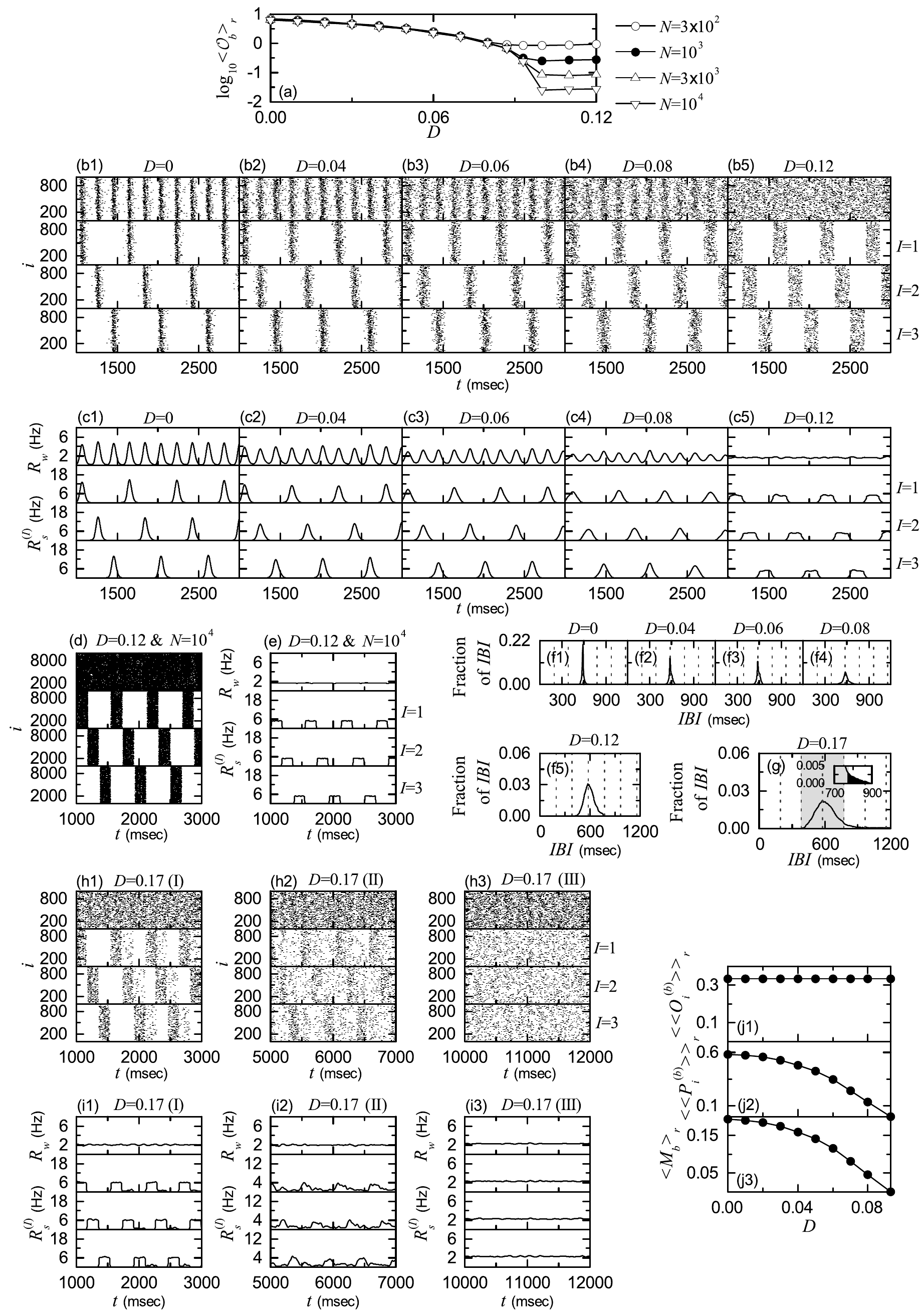}
\caption{
Noise effect in the 1st route for $J_0=3$.
(a) Plots of thermodynamic bursting order parameter $\langle {\cal {O}}_b \rangle_r$  versus the noise intensity $D$.
(b1)-(b5) Raster plots of burst onset times in the whole population and in the $I$ th cluster ($I=1$, 2, and 3) for various values of $D$. (c1)-(c5) IWPBRs $R_w(t)$ of the whole population and ISPBRs $R^{(I)}_s(t)$ of the $I$th cluster ($I=1,$ 2, and 3) for various values of $D$. $D=0.12$ and $N=10^4$: (d) raster plots of burst onset times in the whole population and in the $I$ th cluster ($I=1$, 2, and 3) and (e) IWPBR $R_w(t)$ of the whole population and ISPBRs $R^{(I)}_s(t)$ of the $I$th cluster ($I=1,$ 2, and 3). (f1)-(f5) IBI histograms for various values of $D.$ Vertical dotted lines in (f1)-(f4) and in (f5) denote integer multiples of the global period $T_G$ of $R_w(t)$ and the cluster period $T_c$, respectively.
$D=0.17$: (g) delocalized IBI histogram for $D=0.17$, sequential long-term raster plots of burst onset times in the whole population and in the $I$th cluster ($I=$1, 2, and 3) in (h1) the early, (h2) the intermediate, and (h3) the final stages,
and IWPBR $R_w(t)$ of the whole population and ISPBR $R_s^{(I)}(t)$ of the $I$th cluster ($I=$1, 2, and 3) in the (i1) early, (i2) the intermediate, and (i3) the final stages.
Plots of (j1) the average occupation degree $\langle \langle O_i^{(b)} \rangle \rangle_r$, (j2) the average pacing degree $\langle \langle P_i^{(b)} \rangle \rangle_r$, and (j3) the statistical-mechanical bursting measure $\langle M_b \rangle_r$ versus $D$.
}
\label{fig:NE1}
\end{figure}

\subsection{Effects of Noise in The Route $I$: $C-BS~\rightarrow~C-DS~\rightarrow~NC-DS$}
\label{subsec:Route1}
Figure \ref{fig:NE1} shows results on the noise effects in the 1st route $I$ for $J_0=3$. For $D=0$ a 3-cluster burst synchronization ($C-BS$) occurs.
As $D$ is increased and passes a lower threshold $D^*_l~(\simeq 0.093)$, a transition to desynchronization occurs, which may be
described in terms of the bursting order parameter ${\cal{O}}_b$ of Eq.~(\ref{eq:Order}). Figure \ref{fig:NE1}(a) shows a plot of $\log_{10} \langle {\cal{O}}_b \rangle_r$ versus $D$. With increasing $N$, the bursting order parameter $\langle {\cal{O}}_b \rangle_r$ approaches a non-zero limit value for $0 \leq D < D^*_l$, and hence burst synchronization occurs. On the other hand, when passing $D^*_l$ a transition to desynchronization occurs, because $\langle {\cal{O}}_b \rangle_r$ tends to zero, as $N$ is increased. Consequently, for $D > D^*_l$ desynchronized states appear due to a destructive role of noise to spoil the burst synchronization.

This kind of transition from 3-cluster burst synchronization to 3-cluster desynchronization ($C-DS$) may also be well seen in the raster plots of burst onset times in the whole population and in the $I$th clusters $(I=$1, 2, and 3).
Figures \ref{fig:NE1}(b1)-\ref{fig:NE1}(b5) show such raster plots for $D=$0, 0.04, 0.06, 0.08, and 0.12, respectively. Their corresponding IWPBR $R_w(t)$ and the ISPBR $R_s^{(I)}(t)$ are also given in Figs.~\ref{fig:NE1}(c1)-\ref{fig:NE1}(c5) when $D=$0, 0.04, 0.06, 0.08, and 0.12, respectively. For $D=0$, bursting stripes (representing burst synchronization) appear successively in the raster plot in the whole population [see the top panel of Fig.~\ref{fig:NE1}(b1)], and the corresponding IWPBR $R_w(t)$ exhibits a slow-wave oscillation with the whole-population frequency $f_b^{(w)}~(\simeq 5.2$ Hz), as shown in the top panel of Fig.~\ref{fig:NE1}(c1). The whole population is segregated into 3 clusters ($I=$1, 2, and 3), which is well seen in the raster plots for the clusters [see the $I=1$, 2, and 3 panels in Fig.~\ref{fig:NE1}(b1)]. We note that bursting stripes in each cluster appear successively every 3rd cycle of $R_w(t)$, and the corresponding ISPBR $R_s^{(I)}(t)$ exhibits a regular oscillation with the sub-population frequency $f_b^{(I)}~(\simeq {\frac {f_b^{(w)}}{3}})$. In this way, 3-cluster burst synchronization appears for $D=0$. In this case, a single peak appears at $3~T_G$ [$T_G~(\simeq 193.4$ msec): global period of $R_w(t)$] in the IBI histogram, as shown in Fig.~\ref{fig:NE1}(f1).

As $D$ is increased from 0, the 3-cluster burst synchronization for $D=0$ persists, but its degree becomes more and more worse due to a destructive role of noise to spoil the burst synchronization. As shown in Figs.~\ref{fig:NE1}(b2)-\ref{fig:NE1}(b4), with increasing $D$, bursting stripes in the whole population and in each $I$th
($I=1$, 2, and 3) cluster become smeared more and more. Hence, amplitudes of $R_w(t)$ and $R_s^{(I)}(t)$ also decrease, as $D$ is increased [see Figs.~\ref{fig:NE1}(c2)-\ref{fig:NE1}(c4)]. Peaks in the IBI histograms also become broader (along with decrease in their heights), with increasing $D$ [see Figs.~\ref{fig:NE1}(f2)-\ref{fig:NE1}(f4)].

Eventually, when passing a lower threshold $D^*_l~(\simeq 0.093)$, a transition to 3-cluster desynchronization occurs.
Consequently, desynchronized 3-cluster states appear for $D>D^*_l$. As an example, see the raster plots in Fig.~\ref{fig:NE1}(b5) and the IWPBR $R_w(t)$ and the ISPBR $R_s^{(I)}(t)$ in Fig.~\ref{fig:NE1}(c5) for $D=0.12$.
Burst onset times in the raster plot in the whole population seem to be completely scattered, and the corresponding
IWPBR $R_w(t)$ is nearly stationary.
However, we note that, for $D=0.12$ bursting bands in the raster plot in each cluster are preserved (i.e., 3-clusters are preserved). For each cluster, burst onset times in each bursting band are nearly completely scattered (i.e., nearly desynchronized), and hence a square-wave-like oscillation occurs in each ISPBR $R_s^{(I)}(t)$. During the ``silent'' part (without burstings) for about $\frac {2P} {3}$, $R_s^{(I)}(t)=0$ (which corresponds to the bottom part), while in the bursting band for about $\frac {P} {3}$, $R_s^{(I)}(t)$ rapidly increases to the nearly flat top, and then decreases rapidly; $P~(\simeq 572.5$ msec) corresponds to the average period of the square-wave oscillation. Through repetition of this process $R_s^{(I)}(t)$ exhibits a square-wave-like oscillation. In this case, the IBI histogram in Fig.~\ref{fig:NE1}(f5) becomes broader in comparison with those in the cases of burst synchronization, and its peak appears at $T_{peak} \simeq 572.5$ msec (corresponding to the period $P$ of the square-wave oscillation).

To examine the square-wave-like behavior more clearly, the number of HR neurons is increased from $N=10^3$ to $10^4$. In this case, raster plots in the whole population and the clusters and their corresponding IWPBR $R_w(t)$ and the ISPBR
$R_s^{(I)}(t)$ are shown in Figs.~\ref{fig:NE1}(d) and \ref{fig:NE1}(e), respectively. For the whole population, burst onset times are more completely scattered, and hence the corresponding IWPBR $R_w(t)$ is more stationary. Moreover, for each cluster bursting bands in the raster plot show clearly the clustering structure, and hence the corresponding ISPBR
$R_s^{(I)}(t)$ shows square-wave oscillations more clearly. Thus, for each cluster burst onset times in bursting bands are completely scattered, and they show a desynchronized state. In this way, 3-cluster desynchronization appears, as $D$ passes $D^*_l$.

However, as $D$ is further increased and passes a higher threshold $D^*_h~(\simeq 0.15)$, clusters are broken up via intercluster hoppings due to another destructive role of noise to break up the clusters.
Hence, for $D>D^*_h$ non-cluster desynchronized states appear. As an example, we consider the case of $D=0.17$.
In this case, the IBI histogram is shown in Fig.~\ref{fig:NE1}(g). Its peak is located at $T_{peak}~(\simeq$ 572.5 msec).
We note that some fraction of IBIs with larger than $4~T_c$ ($T_c:$ cluster period corresponding to $T_{peak}/3$) appear
(i.e., late burstings occur), as clearly shown in the inset of Fig.~\ref{fig:NE1}(g). Thus, delocalization of IBIs occurs by crossing the right boundary (corresponding to $4~T_c$), which is in contrast to all the cases of cluster burst synchronization where IBIs are localized in a range of $2~T_c < IBI < 4~T_c$ [see Figs.~\ref{fig:NE1}(f1)-\ref{fig:NE1}(f5)].

Due to  appearance of delocalized IBIs larger than $4~T_c$ (i.e., because of occurrence of late burstings), forward intercluster hoppings between the 3 clusters occur, which leads to break-up of 3 clusters.
Similar to the case of $J_0=0.13$ in the absence of noise ($D=0$) [see Figs.~\ref{fig:NC1}(b1)-\ref{fig:NC1}(b3)],
intercluster hoppings between the 3 clusters may be well seen in sequential long-term raster plots of burst onset times in the whole population and in the $I$th ($I=1$, 2, and 3) clusters. Figures \ref{fig:NE1}(h1)-\ref{fig:NE1}(h3) show such raster plots, corresponding to (h1) the early, (h2) the intermediate, and (h3) the final stages.
As the time $t$ is increased, forward intercluster hoppings occur in a cyclic way [$I$ $\rightarrow$ $I+1$ $\rightarrow$ $I+2$  $\rightarrow$ $I$] due to occurrence of late burstings.
In the final stage after a sufficiently long time, intercluster hoppings between clusters are more and more intensified, which leads to complete break-up of clusters. Consequently, burst onset times in the raster plots are completely scattered in a nearly uniform way, independently of $I=1,$ 2, and 3, as shown in Fig.~\ref{fig:NE1}(h3).
Figures \ref{fig:NE1}(i1)-\ref{fig:NE1}(i3) also show the IWPBR $R_w(t)$ and the ISPBR $R_s^{(I)}(t)$ ($I$=1, 2, and 3), corresponding to the above raster plots in Figs.~\ref{fig:NE1}(h1)-\ref{fig:NE1}(h3).
With increase in the time $t$, the initial square-wave oscillations in $R_s^{(I)}(t)$ are transformed into nearly stationary
ones, independently of $I$ [see the final stage in Fig.~\ref{fig:NE1}(i3)]. Thus, non-cluster desynchronized state
appears for $D=0.17$.

Figures \ref{fig:NE1}(j1)-\ref{fig:NE1}(j3) show the average occupation degree $\langle \langle O_i^{(b)} \rangle \rangle_r$ of Eq.~(\ref{eq:OD}) (representing the average density of bursting stripes), the average pacing degree $\langle \langle P_i^{(b)} \rangle \rangle_r$ of Eq.~(\ref{eq:PD}) (denoting the average degree of phase coherence in bursting stripes), and the statistical-mechanical bursting measure $\langle M_b \rangle_r$ of Eq.~(\ref{eq:BM}) (given by the product of occupation and pacing degrees), respectively, in the range of $0 \leq D < D^*_l$ where 3-cluster burst synchronization occurs.
Obviously, $\langle \langle O_i^{(b)} \rangle \rangle_r = {\frac {1} {3}},$ because 3-clusters persist for $0 \leq D < D^*_l$.
Due to a destructive role of noise to spoil the burst synchronization, as $D$ is increased from 0 to $D^*_l,$
$\langle \langle P_i^{(b)} \rangle \rangle_r$ decreases smoothly from $0.581$ to zero.
Then, the statistical-mechanical bursting measure $\langle M_b \rangle_r$ also makes a smooth decrease from $0.194$ to 0, as in the case of $\langle \langle P_i^{(b)} \rangle \rangle_r$, because $\langle \langle O_i^{(b)} \rangle \rangle_r$ is constant.

\begin{figure}
\includegraphics[width=\columnwidth]{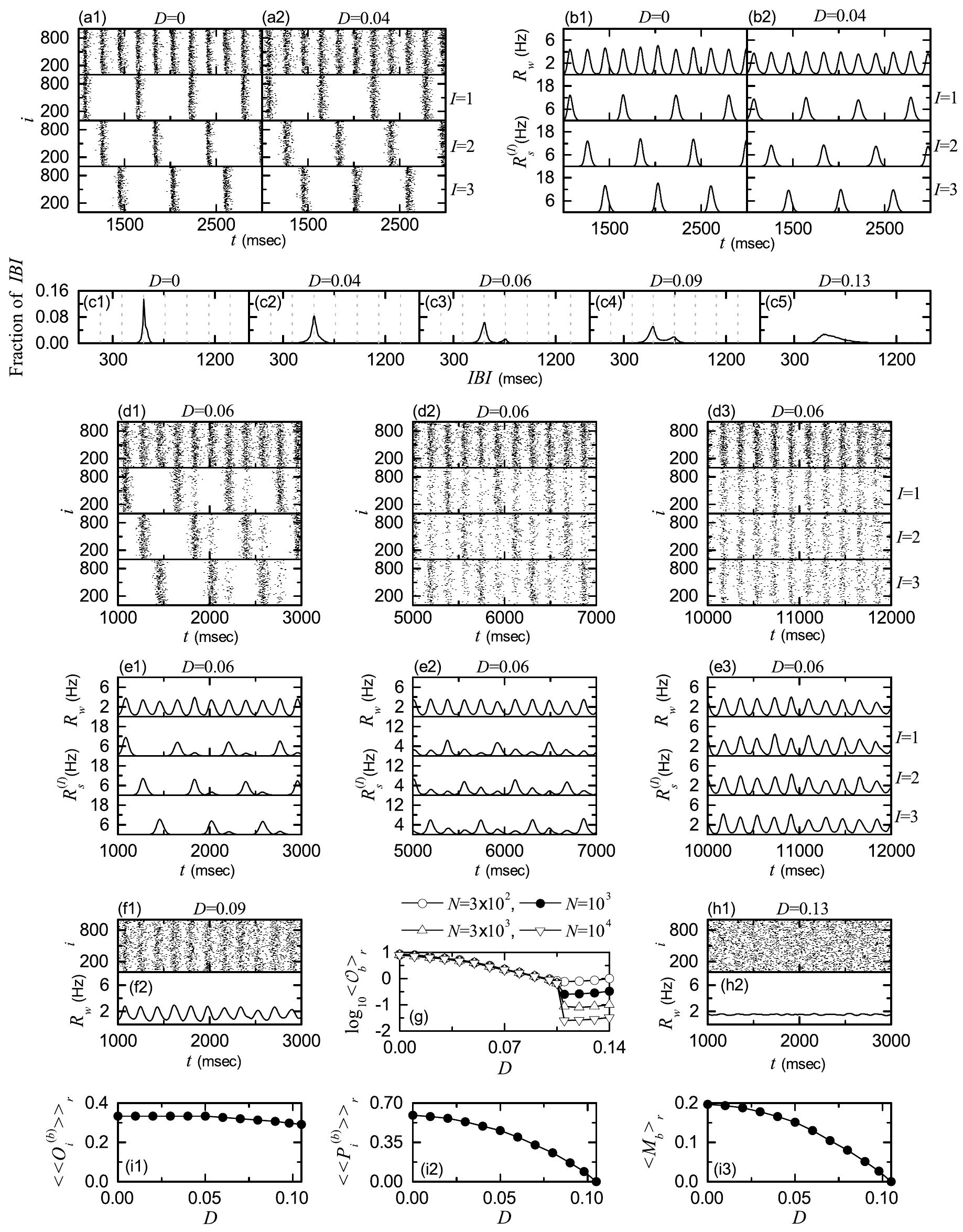}
\caption{
Noise effect in the 2nd route for $J_0=4.5$.
Raster plots of burst onset times in the whole population and in the $I$th cluster ($I=1$, 2, and 3) when $D=$
(a1) 0 and (a2) 0.04. IWPBR $R_w(t)$ of the whole population and ISPBR $R^{(I)}_s(t)$ of the $I$th cluster ($I=1,$ 2, and 3) for $D=$ (b1) 0 and (b2) 0.04. (c1)-(c5) IBI histograms for various values of $D.$ Vertical dotted lines in (c1)-(c4) denote the integer multiples of the global period $T_G$ of $R_w(t)$.
$D=0.06$: sequential long-term raster plots of burst onset times for the whole population and the clusters in (d1) the early, (d2) the intermediate, and (d3) the final stages, and $R_w(t)$ of the whole population and $R_s^{(I)}(t)$ of the clusters in the (e1) early, (e2) the intermediate, and (e3) the final stages.
$D=0.09$: (f1) raster plot of burst onset times in the whole population and (f2) $R_w(t)$ of the whole population.
(g) Plots of thermodynamic bursting order parameter $\langle {\cal {O}}_b \rangle_r$  versus the noise intensity $D$.
For $D=0.13,$ (h1) raster plot of burst onset times in the whole population and (h2) $R_w(t)$ of the whole population.
Plots of (i1) the average occupation degree $\langle \langle O_i^{(b)} \rangle \rangle_r$, (i2) the average pacing degree $\langle \langle P_i^{(b)} \rangle \rangle_r$, and (i3) the statistical-mechanical bursting measure $\langle M_b \rangle_r$ versus $D$.
}
\label{fig:NE2}
\end{figure}

\subsection{Effects of Noise in The Route $II$: $C-BS~\rightarrow~NC-BS~\rightarrow~NC-DS$}
\label{subsec:Route2}
As $J_0$ is increased and passes a threshold $J^{**}~(\simeq 3.7)$, break-up of clusters in the desynchronized states
(e.g., the above case of the route $I$ case for $J_0=3$) no longer occurs. Instead, before a transition to desynchronization, break-up of clusters occurs in the burst-synchronized states. As an example, see the route $II$ for $J_0=4.5$ in Fig.~\ref{fig:SD}. Figure \ref{fig:NE2} shows results on the noise effects in the 2nd route $II$ for $J_0=4.5$. With increasing $D$, noise first breaks up clusters, and then a transition to desynchronization occurs due to another destructive role of noise
to spoil the burst synchronization. Hence, the destructive roles of $D$ is similar to those of $J_0$, shown in Fig.~\ref{fig:ICH} in the absence of noise ($D=0$).

As in the 1st route $I$ for $J_0=3$, when $D=0$ appearance of 3-cluster burst synchronization ($C-BS$) is well shown in the raster plots in the whole population and the clusters ($I=1$, 2, and 3) [see Fig.~\ref{fig:NE2}(a1)] and their corresponding IWPBR $R_w(t)$ and the ISPBR $R_s^{(I)}(t)$ [see Fig.~\ref{fig:NE2}(b1)]. For each cluster, bursting stripes appear every 3rd cycle of $R_w(t)$, which results in emergence of 3-cluster burst synchronization. In this case, the IBI histogram has a single peak at $3~T_G$ [$T_G:$ global period of $R_w(t)$], as shown in Fig.~\ref{fig:NE2}(c1).
As $D$ is increased, bursting stripes in the raster plots become smeared, due to a destructive role of noise to spoil the burst synchronization, and hence amplitudes of $R_w(t)$ and $R_s^{(I)}(t)$ are decreased [e.g., see Figs.~\ref{fig:NE2}(a2) and
\ref{fig:NE2}(b2) for $D=0.04$]. In this case, the IBI histogram has a broad single peak with lower height at $3~T_G$, as shown in Fig.~\ref{fig:NE2}(c2).

Eventually, as $D$ passes a lower threshold $D^{**}_l~(\simeq 0.05)$, the IBI histogram begins to have a new minor peak at
$4~T_G,$ in addition to the major peak at $3~T_G$, as shown in Fig.~\ref{fig:NE2}(c3) for $D=0.06$. Hence, individual HR neurons begin to exhibit burstings intermittently at a 4th cycle of $R_w(t)$ via burst skipping at its 3rd cycle. Due to occurrence of late burstings via burst skippings, clusters become broken up via forward intercluster hoppings, as in the case of Fig.~\ref{fig:ICH} for $J_0=10$ in the absence of noise ($D=0$). As a result, non-cluster burst synchronization ($NC-BS$) without dynamical clusterings appears in the whole population. As an example, we consider the case of $D=0.06$. Similar to the case in Figs.~\ref{fig:ICH}(c1)-\ref{fig:ICH}(c3) and Figs.~\ref{fig:ICH}(d1)-\ref{fig:ICH}(d3), intercluster hoppings for $D=0.06$ are well seen in sequential long-term raster plots of burst onset times in the whole population and in the $I$th ($I=1$, 2, and 3) clusters [see Figs.~\ref{fig:NE2}(d1)-\ref{fig:NE2}(d3)] and in the corresponding IWPBR $R_w(t)$ of the whole population and the ISPBR $R_s^{(I)}(t)$ of the clusters [see Figs.~\ref{fig:NE2}(e1)-\ref{fig:NE2}(e3)].
Here, Figs.~\ref{fig:NE2}(d1) and \ref{fig:NE2}(e1), Figs.~\ref{fig:NE2}(d2) and \ref{fig:NE2}(e2), and Figs.~\ref{fig:NE2}(d3) and \ref{fig:NE2}(e3) show the initial, the intermediate, and the final stages, respectively. With increasing the stage, intercluster hoppings are more and more intensified due to burst skippings, which results in complete break-up of clusters. Thus, after a sufficiently long time, raster plots in the clusters ($I=1$, 2, and 3) are essentially the same, irrespectively of $I$. Although clusters are broken up, bursting stripes persist, and hence burst synchronization without dynamical clusterings occurs in the whole population.

With increasing $D$ from 0.06, the degree of burst synchronization is decreased due to a destructive role of noise to
spoil the burst synchronization. In the IBI histogram for $D=0.09$, the height of the peak at $3~T_G$ is decreased, while
the height of the peak at $4~T_G$ increases a little [see Fig.~\ref{fig:NE2}(c4)]. Thus, the IBI histogram becomes broader,
and burst skippings are enhanced. Consequently, intercluster hoppings are more intensified.
Figures \ref{fig:NE2}(f1) and \ref{fig:NE2}(f2) show the raster plot and the corresponding IWPBR $R_w(t)$
for $D=0.09$, respectively. In comparison with the case of $D=0.06$, bursting stripes are more smeared and amplitudes of $R_w(t)$ are decreased. Eventually, when passing a higher threshold $D^{**}_h~(\simeq 0.105)$, a transition from non-cluster burst synchronization to desynchronization ($NC-DS$) occurs.

The bursting order parameter ${\cal{O}}_b$ of Eq.~(\ref{eq:Order}) may describe well a transition from burst synchronization to desynchronization. Figure \ref{fig:NE2}(g) shows a plot of $\log_{10} \langle {\cal{O}}_b \rangle_r$ versus $D$.
As $N$ is increased, the bursting order parameter $\langle {\cal{O}}_b \rangle_r$ approaches a non-zero limit value for $0 \leq D < D^{**}_h~(\simeq 0.105)$, and hence burst synchronization occurs. In contrast, when passing $D^{**}_h$ a transition to desynchronization occurs, because $\langle {\cal{O}}_b \rangle_r$ tends to zero, with increasing $N$.
Consequently, for $D > D^{**}_h$ desynchronized states appear due to a destructive role of noise to spoil the burst synchronization. As an example of desynchronized state, we consider the case of $D=0.13$. With increasing $D$ the two peaks in the IBI histogram for $D=0.09$ are merged, and then it has a broad single maximum, as shown in Fig.~\ref{fig:NE2}(c5) for $D=0.13$. In this case, burst onset times are completely scattered in the raster plot, and the corresponding IWPBR $R_w(t)$ is nearly stationary [see Figs.~\ref{fig:NE2}(h1) and \ref{fig:NE2}(h2)].

Figures \ref{fig:NE2}(i1)-\ref{fig:NE2}(i3) show the average occupation degree $\langle \langle O_i^{(b)} \rangle \rangle_r$   (denoting the average density of bursting stripes), the average pacing degree $\langle \langle P_i^{(b)} \rangle \rangle_r$   (representing the average degree of phase coherence in bursting stripes), and the statistical-mechanical bursting measure $\langle M_b \rangle_r$ (given by the product of occupation and pacing degrees), respectively, in the range of $0 \leq D < D^{**}_h$ where burst synchronization occurs. In the range of $0 \leq D < D^{**}_l$, 3-cluster burst synchronization appears, and hence $\langle \langle O_i^{(b)} \rangle \rangle_r = {\frac {1} {3}}.$ However, as a result of break-up of clusters, for $D>D^{**}_l,$ $\langle \langle O_i^{(b)} \rangle \rangle_r$ decreases slowly to a limit value ($\simeq 0.291)$ for $D=D^{**}_h$, due to bursting skippings. With increasing $D$ from 0 to $D^{**}_h,$ bursting stripes become more and more smeared due to a destructive role of noise to spoil the burst synchronization, and eventually they become completely overlapped for $D=D^{**}_h$.
Hence, in the range of $0 \leq D < D^{**}_h$, $\langle \langle P_i^{(b)} \rangle \rangle_r$ decreases smoothly from $0.587$ to zero. Then, through product of the occupation and pacing degrees of burst onset times, the statistical-mechanical bursting measure $\langle M_b \rangle_r$ also makes a smooth decrease from $0.196$ to 0, like the case of $\langle \langle P_i^{(b)} \rangle \rangle_r$, because variations in $\langle \langle O_i^{(b)} \rangle \rangle_r$ are small.

\begin{figure}
\includegraphics[width=\columnwidth]{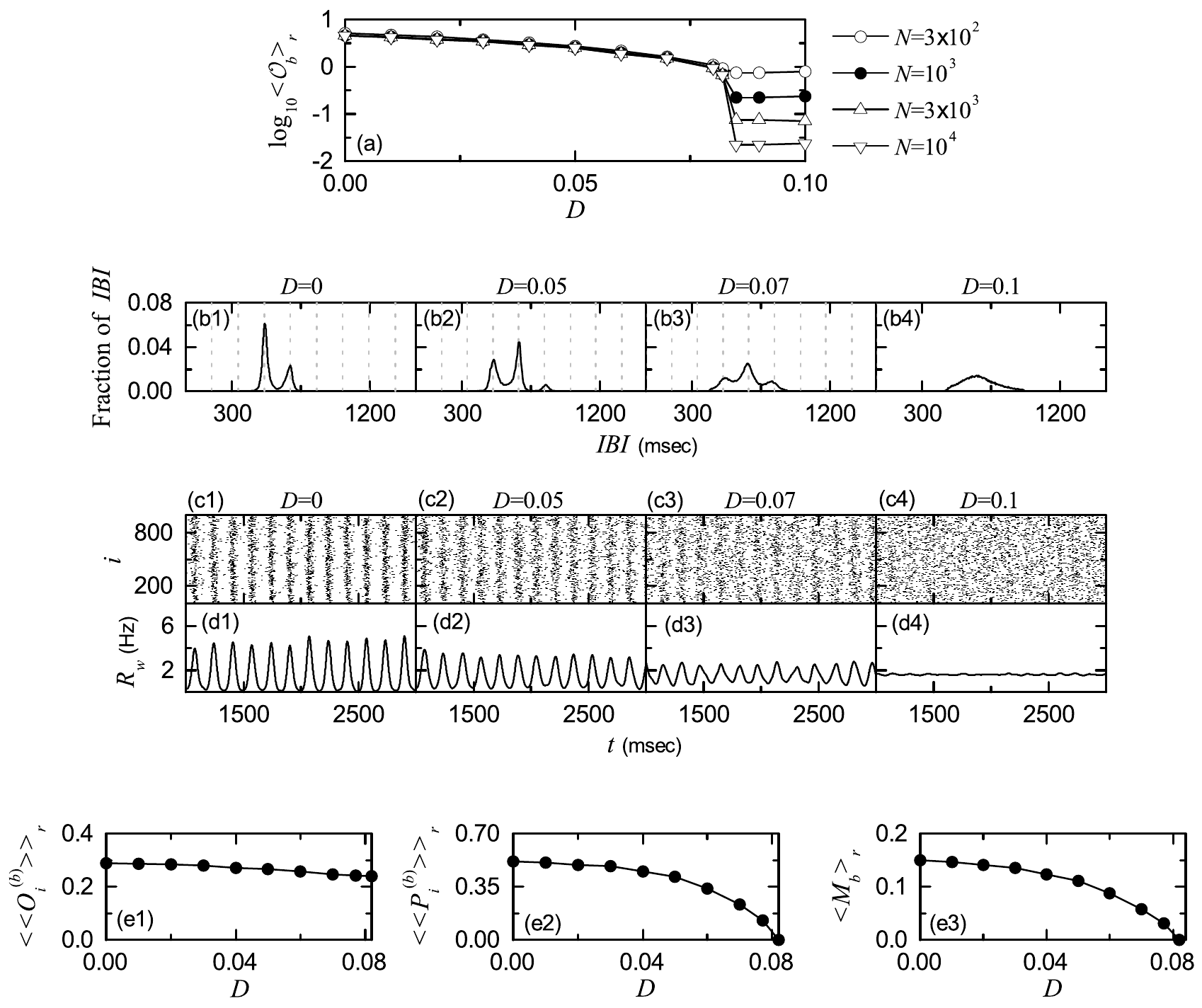}
\caption{
Noise effect in the 3rd route for $J_0=10$.
(a) Plots of thermodynamic bursting order parameter $\langle {\cal {O}}_b \rangle_r$  versus the noise intensity $D$.
IBI histograms for $D=$ (b1) 0, (b2) 0.05, (b3) 0.07, and (b4) 0.1.
Vertical dotted lines in (b1)-(b3) represent integer multiples of the global period $T_G$ of $R_w(t)$.
Raster plots of burst onset times in the whole population when $D=$ (c1) 0, (c2) 0.05, (c3) 0.07, and (c4) 0.1.
IWPBR kernel estimates $R_w(t)$ of the whole population for $D=$ (d1) 0, (d2) 0.05, (d3) 0.07, and (d4) 0.1.
Plots of (e1) the average occupation degree $\langle \langle O_i^{(b)} \rangle \rangle_r$, (e2) the average pacing degree $\langle \langle P_i^{(b)} \rangle \rangle_r$, and (e3) the statistical-mechanical bursting measure $\langle M_b \rangle_r$ versus $D$.
}
\label{fig:NE3}
\end{figure}

\subsection{Effects of Noise in The Route $III$: $NC-BS~\rightarrow~NC-DS$}
\label{subsec:Route3}
Finally, we consider the route $III$ for $J_0=10$ in Fig.~\ref{fig:SD}.
Unlike the above cases of routes $I$ and $II$, in the absence of noise ($D=0$) clusters are broken up due to
burst skippings, and hence non-cluster burst synchronization ($NC-BS$) without dynamical clusterings appears.
In this case, we investigate the noise effect on the non-cluster burst synchronization by increasing $D$, and
due to destructive roles of noise, both intensified intercluster hoppings via burst skippings and smearing of
bursting stripes are thus found.

As shown in the above cases, a transition from burst synchronization to desynchronization may be well described in terms of
the bursting order parameter ${\cal{O}}_b$. Figure \ref{fig:NE3}(a) shows a plot of $\log_{10} \langle {\cal{O}}_b \rangle_r$ versus $D$. With increasing $N$, the bursting order parameter $\langle {\cal{O}}_b \rangle_r$ converges to a non-zero limit value for $0 \leq D < D^{***}~(\simeq 0.082)$. Consequently, burst synchronization occurs.
On the other hand, when passing $D^{***}$ a transition to desynchronization occurs, because $\langle {\cal{O}}_b \rangle_r$ tends to zero, as $N$ is increased. Accordingly, for $D > D^{***}$ desynchronized states ($NC-DS$) appear due to a destructive role of noise to spoil the burst synchronization.

Figures \ref{fig:NE3}(b1)-\ref{fig:NE3}(b4) show the IBI histograms for $D=$0, 0.05, 0.07, and 0.1, respectively.
For $D=0$, a minor peak appears at $4~T_G$, in addition to the major peak at $3~T_G$. Hence, individual HR neurons exhibit burstings intermittently at a 4th cycle of $R_w(t)$ via burst skipping at its 3rd cycle. Due to this type of burst
skippings, intercluster hoppings occur between clusters, and the clusters become broken up. Thus, for $D=0$ non-cluster
burst synchronization without dynamical clusterings appears, in contrast to the above two cases.

With increasing $D$, the height of the peak at $4~T_G$ is increased, while the height of the peak at $3~T_G$ decreases.
Furthermore, a small peak also appears at $5~T_G$, as shown in Fig.~\ref{fig:NE3}(b2) for $D=0.05$. Hence, intercluster
hoppings become intensified due to enhanced burst skippings. With further increase in $D$, these 3 peaks
begin to show a tendency of merging [e.g., see Fig.~\ref{fig:NE3}(b3) for $D=0.07$]. In the desynchronized case of $D=0.1$,
these peaks are completely merged, and then the IBI histogram has a broad single peak.

Figures \ref{fig:NE3}(c1)-\ref{fig:NE3}(c4) show raster plots for $D=0$, 0.05, 0.07, and 0.1, respectively, and
their corresponding IWPBR $R_w(t)$ are also shown in Figs.~\ref{fig:NE3}(d1)-\ref{fig:NE3}(d4), respectively.
As $D$ is increased from 0, bursting stripes in the raster plots become more and more smeared, and amplitudes of $R_w(t)$
also are decreased. Hence, with increasing $D$ the degree of burst synchronization becomes worse, due to a destructive role of
noise to spoil the burst synchronization.

Figures \ref{fig:NE3}(e1)-\ref{fig:NE3}(e3) show the average occupation degree $\langle \langle O_i^{(b)} \rangle \rangle_r$, the average pacing degree $\langle \langle P_i^{(b)} \rangle \rangle_r$, and the statistical-mechanical bursting measure $\langle M_b \rangle_r$  respectively, in the range of $0 \leq D < D^{***}$ where burst synchronization (without dynamical clusterings) occurs. As $D$ is increased from 0 to $D^{***}$, burst skippings become intensified, and hence $\langle \langle O_i^{(b)} \rangle \rangle_r$ decreases smoothly from 0.289 (for $D=0$) to 0.241 (for $D=D^{***}$).
With increasing $D$ from 0 to $D^{***},$ bursting stripes become more and more smeared due to a destructive role of noise to
spoil the burst synchronization, and eventually they become completely overlapped for $D=D^{***}$.
Hence, in the range of $0 \leq D < D^{***}$, $\langle \langle P_i^{(b)} \rangle \rangle_r$ decreases smoothly from $0.516$ to zero. Then, through product of the occupation and pacing degrees of burst onset times, the statistical-mechanical bursting measure $\langle M_b \rangle_r$ also makes a smooth decrease from $0.149$ to 0, as in the case of $\langle \langle P_i^{(b)} \rangle \rangle_r$, because variations in $\langle \langle O_i^{(b)} \rangle \rangle_r$ are small.

\section{Summary and Discussion}
\label{sec:SUM}
We investigated coupling-induced cluster burst synchronization by changing the average coupling strength $J_0$ in an inhibitory Barab\'{a}si-Albert SFN of HR bursting neurons. For sufficiently small $J_0$, non-cluster desynchronized states exist. But,
when passing a critical point $J^*_c~(\simeq 0.16)$, the whole population has been found to be segregated into 3 clusters via a constructive role of synaptic inhibition to stimulate dynamical clusterings between individual burstings, and thus 3-cluster
desynchronized states appear. Our SFN has no internal symmetries, and hence occurrence of clusters in our case has no relation with network topology, in contrast to the case of occurrence of clusters in networks with a certain degree of internal symmetries \cite{Bely}.

We also note that, in the presence of 3 clusters, IBIs of individual HR neurons are localized in a region of $2~T_c < IBI < 4~T_c$ [$T_c:$ cluster period (i.e., average time interval between appearance of successive clusters)].
Thus, we suggest the following criterion, based on the IBI histogram, for emergence of 3-cluster state.
The cluster period $T_c$ is given by $T_{peak}/3$; the peak of the IBI histogram appears at $T_{peak}$.
Localization of IBIs in a region of $2~T_c < IBI < 4~T_c$ leads to occurrence of 3-cluster state.
For $J_0 < J^*_c$, delocalization of IBIs has been found to occur via crossing the right and/or the left boundaries (corresponding to $4~T_c$ and $2~T_c,$ respectively), and thus late and/or early burstings appear.
Through appearance of the late and/or early burstings, forward and/or backward intercluster hoppings have been found to occur, which leads to break-up of the 3 clusters.

As $J_0$ is further increased and passes a lower threshold $J^*_l~(\simeq 0.78$), a transition to 3-cluster burst synchronization has been found to occur due to another constructive role of synaptic inhibition to favor population synchronization. In each cluster, HR neurons make burstings every 3rd cycle of the IWPBR $R_w(t)$.
Therefore, a single peak has been found to appear at $3~T_G$ [$T_G:$ global period of $R_w(t)$] in the IBI histogram; in this case, $T_G = T_c$.  Furthermore, these burstings in each cluster have been found to exhibit burst synchronization. In this way, 3-cluster burst synchronization has been found to emerge. Burst synchronization in the whole population may be well visualized in the raster plot of burst onset times where bursting stripes appear in a regular and successive way, and the corresponding IWPBR $R_w(t)$ shows regular oscillations with the whole-population bursting frequency $f_b^{(w)}$. Moreover, cluster
burst synchronization may also be seen well in the raster plot of burst onset times in each cluster, along with the corresponding ISPBR $R^{(I)}_s(t)$ $(I=$1, 2, and 3) of the sub-populations. Bursting stripes in each cluster appear every 3rd cycle of $R_w(t)$, and the corresponding ISPBR $R_s^{(I)}(t)$ exhibits regular oscillations with the sub-population bursting frequency $f_b^{(I)}~(\simeq {\frac {f_b^{(w)}} {3}})$.

However, with increase in $J_0$ and passing an intermediate threshold $J^*_m~(\simeq 5.2)$, a new peak has been found to appear at $4~T_G$ in the IBI histogram, in addition to the main peak at $3~T_G$. In this case, delocalization of IBIs occurs through crossing the right boundary (corresponding to $4~T_c$), and thus late burstings appear. Hence, HR neurons have been found to exhibit intermittent forward hoppings between the 3 clusters, since they intermittently fire burstings at a 4th cycle of $R_w(t)$ due to burst skipping rather than at its 3rd cycle.
As a result of the intermittent forward intercluster hoppings, the 3 clusters have been found to be integrated into a single one, which was well shown in sequential long-term raster plots of burst onset times. Although the 3 clusters are broken up, burst synchronization has been found to persist in the whole population. As $J_0$ is further increased, forward intercluster hoppings have been found to be intensified due to enhanced burst skippings (e.g., for $J_0 =15$ a 3rd peak appears at $5~T_G$ in the IBI histogram), and bursting stripes have also been found to be smeared more and more because of a destructive role of synaptic inhibition to spoil the burst synchronization. Eventually, as a higher threshold $J^*_h~(\simeq 17.8)$ is passed, a transition to desynchronization has been found to occur. Then, burst onset times are completely scattered in the raster plot due to complete overlap between the bursting stripes, and the IWPBR $R_w(t)$ becomes nearly stationary.

We have also studied the effects of stochastic noise on burst synchronization, and obtained a state diagram in the $J_0-D$ plane. By increasing the noise intensity $D,$ we investigated the noise effects along the 3 routes $I,$ $II,$ and $III$ for $J_0=3$, 4.5, and 10, respectively. For $J^*_l < J_0 < J_m^*$ (where 3-cluster burst synchronization occurs for $D=0$), two cases have been found to appear; the 1st (2nd) case occurs when $J_0 < (>) J^{**} (\simeq 3.7)$. As the 1st example, we considered the 1st route $I$ for $J_0=3$. With increasing $D$, bursting stripes become just smeared (without intercluster hoppings) due to a destructive role of noise to spoil the cluster burst synchronization. Eventually when passing a lower threshold $D^*_l~(\simeq 0.093)$, a transition from the 3-cluster burst synchronization to desynchronization has been found to occur via complete overlap between the bursting stripes. As a result, desynchronized 3-cluster states appear for $D>D^*_l$.
In the presence of 3 clusters, IBIs have been found to be localized in a range of $2~T_c < IBI < 4~T_c$, independently of whether they are synchronized or desynchronized. However, as $D$ is further increased and passes a higher threshold $D^*_h ~(\simeq 0.15)$, delocalization has been found to occur via crossing the right boundary (corresponding to $4~T_c$), and thus late burstings appear. Due to appearance of such late burstings, forward intercluster hoppings have been found to occur between the 3 clusters, which results in break-up of the 3 clusters. As a result, (non-cluster) desynchronized states without dynamical clusterings appear for $D>D^*_h$.

On the other hand, in the 2nd route $II$ for $J_0=4.5$, intercluster hoppings have been found to occur before desynchronization when passing a lower threshold $D_l^{**}(\simeq 0.05)$, in contrast to the case of the 1st route.
For $D>D_l^{**}$, delocalization of IBIs has been found to occur because the IBI histogram has a new minor peak at $4~T_G,$ in addition to the major peak at $3~T_G$. In this case, individual HR neurons exhibit burstings intermittently at a 4th cycle of $R_w(t)$ via burst skipping at its 3rd cycle. Due to occurrence of late burstings via burst skippings, clusters has been found to become broken up via forward intercluster hoppings between the 3 clusters, as in the case of $J_0=10$ in the absence of noise ($D=0$). As a result, non-cluster burst synchronization without dynamical clusterings persists in the whole population, in contrast to the above 1st example. Then, a transition to (non-cluster) desynchronization has also been found to occur when passing a higher threshold $D_h^{**}(\simeq 0.105)$, due to a destructive role of noise to spoil the burst synchronization.
As a 3rd example, we considered the 3rd route $III$ for $J_0=10$ (where (non-cluster) burst synchronization without dynamical clusterings exists for $D=0$). With increasing $D$ from 0, both smearing and intercluster hoppings have been found to be intensified due to a destructive role of noise, and when passing a threshold $D^{***}~(\simeq 0.082)$, (non-cluster) desynchronized states have been found to occur.

As shown in these 3 examples, the stochastic noise plays destructive dual roles to spoil the burst synchronization and to break up  clusters. We also note that, in the present work in a population of (self-firing) suprathreshold bursting neurons, noise makes just destructive effects on population states without showing any constructive role. These noise effects in the suprathreshold case are in contrast to those in previous works \cite{Kim1,SBS} on stochastic burst synchronization (SBS) in a population of (non-self-firing) subthreshold bursting neurons where SBS was found to appear in an intermediate range of noise intensity via competition between the constructive and the destructive roles of noise.

As a complex network, we also considered another Watts-Strogatz small-world network of inhibitory HR neurons \cite{SWN}, and found emergence of cluster burst synchronization, as in the case of SFN. Hence, this kind of cluster burst synchronization seems to occur, independently of network architecture. In addition to the HR model of spike-driven burstings, we considered the Plant model of slow-wave burstings \cite{Longtin,Plant}. In the SFN of inhibitory Plant neurons, 2-cluster burst synchronization has also been found to occur. The number of clusters varies depending on the type of individual burstings. In the case of spike-driven burstings for the HR neurons, rapid hyperpolarization follows the active bursting phase of repetitive spikes [see
Fig.~\ref{fig:Single}(c)], and hence nearly whole silent phase may become available for burstings of HR neurons belonging to the other two clusters. On the other hand, in the case of slow-wave burstings for the Plant neurons, hyperpolarization occurs near the middle of the silent phase (see Fig.~1(a) in \cite{Longtin}), and thus burstings belonging to only one additional cluster may occur during the silent phase. In this way, occurrence of cluster burst synchronization in inhibitory networks seems to be generic, independently of types of constituent bursting neurons, although the number of clusters depend on specific types of individual burstings. We also considered an SFN of excitatory HR neurons. In the case of phase-attractive synaptic excitation, we found only full synchronization (i.e., all bursting neurons exhibit burstings in each bursting stripes) without any dynamical clusterings, in contrast to the case of phase-repulsive synaptic inhibition which is an essential factor for emergence of clusters.

Finally, we expect that our results on burst synchronization, associated with neural information processes in health and disease, would make some contributions for understanding mechanisms of emergence and break-up of cluster burst synchronization and effects of stochastic noise on burst synchronization.

\section*{Acknowledgments}
This research was supported by the Basic Science Research Program through the National Research Foundation of Korea (NRF) funded by the Ministry of Education (Grant No. 20162007688).


\begin{thebibliography}{}
\bibitem{PIR1} D. Golomb and J. Rinzel, Physica D {\bf 72}, 259 (1994).
\bibitem{BSsync1} R. C. Elson, A. I. Selverston, R. Huerta, N. F. Rulkov, M. I. Rabinovich, and H. D. I. Abarbanel, Phys. Rev. Lett. {\bf 81}, 5691 (1998).
\bibitem{BSsync2} E. A. Stern, D. Jaeger, and C. J. Wilson, Nature {\bf 394}, 475 (1998).
\bibitem{BSsync3} P. Varona, J. J. Torres, H. D. I. Abarbanel, M. I. Rabinovych, and R. C. Elson, Biol. Cybern. {\bf 84}, 91 (2001).
\bibitem{BSsync4} C. van Vreeswijk and D. Hansel, Neural Comput. {\bf 13}, 959 (2001).
\bibitem{BSsync5} M. Dhamala, V. Jirsa, and M. Ding, Phys. Rev. Lett. {\bf 92}, 028101 (2004).
\bibitem{BSsync6} M. V. Ivanchenko, G. Osipov, V. Shalfeev, and J. Kurths, Phys. Rev. Lett. {\bf 93}, 134101 (2004).
\bibitem{PIR2} D. T. W. Chik, S. Coombes, and Z. D. Wang, Phys. Rev. E {\bf 70}, 011908 (2004).
\bibitem{BSsync7} A. Shilnikov and G. Cymbalyuk, Phys. Rev. Lett. {\bf 94}, 048101 (2005).
\bibitem{BSsync8} X. Shi and Q. Lu, Chinese Phys. {\bf 14}, 77 (2005).
\bibitem{BSsync9} G. Tanaka, B. Ibarz, M.A. Sanjuan, and K. Aihara, Chaos {\bf 16}, 013113 (2006).
\bibitem{BSsync10} T. Pereira, M. Baptista, and J. Kurths, Eur. Phys. J. Spec. Top. {\bf 146}, 155 (2007).
\bibitem{BSsync11} C. A. S. Batista, A.M. Batista, J. A. C. de Pontes, R. L. Viana, and S. R. Lopes, Phys. Rev. E {\bf 76}, 016218 (2007).
\bibitem{BSsync12} C. A. S. Batista, A. M. Batista, J. C. A. de Pontes, S. R. Lopes, and R. L. Viana, Chaos Soliton. Fract. {\bf 41}, 2220 (2009).
\bibitem{BSsync13} X. Shi and Q. Lu, Physica A {\bf 388}, 2410 (2009).
\bibitem{BSsync14} Q. Wang, M. Perc, Z. Duan, and G. Chen, Phys. Rev. E {\bf 80}, 026206 (2009).
\bibitem{BSsync15} C. A. S. Batisa, S. R. Lopes, R. L. Viana, and A. M. Batisa, Neural Netw. {\bf 23}, 114 (2010).
\bibitem{BSsync16} X. Sun, J. Lei, M. Perc, J. Kurths, and G. Chen, Chaos {\bf 21}, 016110 (2011).
\bibitem{BSsync17} H. Yu, J. Wang, B. Deng, X. Wei, Y. K. Wong, W. L. Chan, K.M. Tsang, and Z. Yu, Chaos {\bf 21}, 013127 (2011).
\bibitem{BSsync18} Q.-Y. Wang, A. Murks, M. Perc, and Q.-S. Lu, Chinese Phys. B {\bf 20}, 040504 (2011).
\bibitem{BSsync19} Q. Wang, G. Chen, and M. Perc, PLoS ONE {\bf 6}, e15851 (2011).
\bibitem{BSsync20} C. A. Batista, E. L. Lameu, A. M. Batista, S. R. Lopes, T. Pereira, G. Zamora-Lopez, J. Kurths, and R. L. Viana, Phys. Rev. E {\bf 86}, 016211 (2012).
\bibitem{BSsync21} E. L. Lameu, C. A. S. Batista, A. M. Batista, K. Larosz, R. L. Viana, S. R. Lopes, and J. Kurths, Chaos {\bf 22}, 043149 (2012).
\bibitem{PIR3} A. J. Langdon, M. Breakspear, and S. Coombes, Phys. Rev. E {\bf 86}, 061903 (2012).
\bibitem{BSsync22} L. Duan, D. Fan, and Q. Lu, Cogn. Neurodyn. {\bf 7}, 341 (2013).
\bibitem{BSsync23} P. Meng, Q. Wang, and Q. Lu, Cogn. Neurodyn. {\bf 7}, 197 (2013).
\bibitem{BSsync24} H. Wang, Q. Wang, Q. Lu, and Y. Zheng, Cogn. Neurodyn. {\bf 7}, 121 (2013).
\bibitem{BSsync25} T. de L. Prado, S. R. Lopes, C. A. S. Batista, J. Kurths, and R. L. Viana, Phys. Rev. E {\bf 90}, 032818 (2014).
\bibitem{BSsync26} B. A. S. Ferrari, R. L. Viana, S. R. Lopes, and R. Stoop, Neural Netw. {\bf 66}, 107 (2015).
\bibitem{Kim1} S.-Y. Kim and W. Lim, Cogn. Neurodyn. {\bf 9}, 179 (2015).
\bibitem{Kim2} S.-Y. Kim and W. Lim, Physica A {\bf 438}, 544 (2015).
\bibitem{NN-SFN} S.-Y. Kim and W. Lim, Neural Netw. {\bf 79}, 53 (2016).
\bibitem{SBS} S.-Y. Kim and W. Lim, Cogn. Neurodyn. {\bf 12}, 315 (2018).
\bibitem{Burst2} E. M. Izhikevich, Scholarpedia {\bf {1(3)}}, 1300 (2006).
\bibitem{Izhi} E. M. Izhikevich, Int. J. Bifurcat.  Chaos {\bf 10}, 1171 (2000).
\bibitem{Burst1} {\it Bursting: The Genesis of Rhythm in the Nervous System,} edited by S. Coombes and P. C. Bressloff (World Scientific, Singapore, 2005).
\bibitem{Rinzel1} J. Rinzel, in {\em Ordinary and Partial Differential Equations}, edited by B. D. Sleeman and R. J. Jarvis, Lecture Notes in Mathematics Vol. 1151 (Springer, Berlin, 1985), pp.~304-316.
\bibitem{Rinzel2} J. Rinzel, in {\em Mathematical Topics in Population Biology, Morphogenesis, and Neurosciences}, edited by E. Teramoto and M. Yamaguti, Lecture Notes in Biomathematics Vol. 71 (Springer, Berlin, 1987), pp.~267-281.
\bibitem{Burst3} E. M. Izhikevich, {\it Dynamical Systems in Neuroscience} (MIT Press, Cambridge, 2007).
\bibitem{Izhi2} E. M. Izhikevich, IEEE Trans. Neural Netw. {\bf 15}, 1063 (2004).
\bibitem{Burst4} R. Krahe and F. Gabbian, Nat. Rev. Neurosci. {\bf 5}, 13 (2004).
\bibitem{Burst5} J. Lisman, Trends Neurosci. {\bf 20}, 38 (1997).
\bibitem{Burst6} E. N. Izhikevich, N. S. Desai, E. C. Walcott, and F. C. Hoppensteadt, Trends Neurosci. {\bf 26}, 161 (2003).
\bibitem{CT1} B. W. Connors and M. J. Gutnick, Trends Neurosci. {\bf 13}, 99 (1990).
\bibitem{CT2} C. M. Gray and D. A. McCormick, Science {\bf 274}, 109 (1996).
\bibitem{TRN1} R. L. Llin\'{a}s and H. Jahnsen, Nature {\bf 297}, 406 (1982).
\bibitem{TRN2} D. A. McCormick and J. R. Huguenard, J. Neurophysiol. {\bf 8}, 1384 (1992).
\bibitem{TR} S. H. Lee, G. Govindaiah, and C. L. Cox, J. Physiol. {\bf 582}, 195 (2007).
\bibitem{HP} H. Su, G. Alroy, E. D. Kirson, and Y. Yaari, J. Neurosci. {\bf 21}, 4173 (2001).
\bibitem{PC} M. D. Womack and K. Khodakhah, J. Neurosci. {\bf 22}, 10603 (2002).
\bibitem{PBC1} T. R. Chay and J. Keizer, Biophys. J. {\bf 42}, 181 (1983).
\bibitem{PBC2} T. A. Kinard, G. de Vries, and A. Sherman, Biophys. J. {\bf 76}, 1423 (1999).
\bibitem{PBC3} M. Pernarowski, R. M. Miura, and J. Kevorkian, SIAM J. Appl. Math. {\bf 52}, 1627 (1992).
\bibitem{BC1} C. A. Del Negro, C.-F. Hsiao, S. H. Chandler, and A. Garfinkel, Biophys. J. {\bf 75}, 174 (1998).
\bibitem{BC2} R. J. Butera, J. Rinzel, and J. C. Smith, J. Neurophysiol. {\bf 82}, 382 (1999).
\bibitem{Spindle1} M. Steriade, D. A. McCormick, and T. J. Sejnowski, Science {\bf 262}, 679 (1993).
\bibitem{Spindle2} M. Bazhenov and I. Timofevv, Scholarpedia {\bf {1(6)}}, 1319 (2006).
\bibitem{Spindle3} S. Gais, W. Plihal, U. Wagner, and J. Born, Nat. Neurosci. {\bf 3}, 1335 (2000).
\bibitem{Spindle4} T. J. Sejnowski and A. Destexhe, Brain Res. {\bf 886}, 208 (2000).
\bibitem{PD1} M. Bevan, P. Magill, D. Terman, J. Bolam, and C. Wilson, Trends Neurosci. {\bf 25}, 525 (2002).
\bibitem{PD2} P. Brown, Cur. Opin. Neurobiol. {\bf 17}, 656 (2007).
\bibitem{PD3} C. Park, R. M. Worth, and L. L. Rubchinsky, J. Neurophysiol. {\bf 103}, 2703 (2010).
\bibitem{PD4} C. Hammond, H. Bergman, and P. Brown, Trends Neurosci. {\bf 30}, 357 (2007).
\bibitem{PD5} P. J. Uhlhaas and W. Singer, Neuron {\bf 52}, 155 (2006).
\bibitem{Epilepsy} R. Fisher, W. van Emde Boas, W. Blume, C. Elger, P. Genton, P. Lee, and J. Engel, Epilepsia {\bf 46}, 470 (2005).
\bibitem{CS1} V. N. Belykh, G. V. Osipov, V. S. Petrov, J. A. K. Suykens, and J. Vandewalle, Chaos {\bf 18}, 037106 (2008).
\bibitem{CS2} S. J. Moon, K. A. Cook, K. Rajendran, K. G. Kevrekidis, J. Cisternas, and C. R. Liang, J. Math. Neurosci.
{\bf 5}, 2 (2015).
\bibitem{Josep1} D. G. Aronson, M. Golubitsky, and M. Krupa, Nonlinearity {\bf 4}, 861 (1991).
\bibitem{Josep2} K. Wisenfeld, P. Colet, and S. Wisenfeld, Phys. Rev. Lett. {\bf 76}, 404 (1996).
\bibitem{CSExp1} I. Z. Kiss, Y. Zhai, and H. Hudson, Phys. Rev. Lett. {\bf 94}, 248301 (2005).
\bibitem{CSExp2} A. F. Taylor, P. Kapetanopoulos, B. J. Whitaker, R. Toth, L. Bull, and M. R. Tinsley, Phys. Rev. Lett. {\bf 100},
214101 (2008).
\bibitem{CSExp3} K. Miyakawa, T. Okano, and S. Yamazaki, J. Phy. Soc. Japan {\bf 82}, 034005 (2013).
\bibitem{Genetic} J. Zhang, Z. Yuan, and T. Zhou, Phys. Rev. E {\bf 79}, 041903 (2009).
\bibitem{Sporns} O. Sporns, {\it Networks of the Brain} (MIT Press, Cambridge, 2011).
\bibitem{Buz2} G. Buzs$\acute{\rm a}$ki, C. Geisler, D. A. Henze, and X.-J. Wang, Trends Neurosci. {\bf 27}, 186 (2004).
\bibitem{CN1} D. B. Chklovskii, B. W. Mel, and K. Svoboda, Nature {\bf 431}, 782 (2004).
\bibitem{CN2} S. Song, P. J. Sj$\ddot{\rm o}$str$\ddot{\rm o}$m, M. Reigl, S. Nelson, and D. B. Chklovskii, PLoS Biol. {\bf 3}, e68 (2005).
\bibitem{CN3} O. Sporns and C. J. Honey, Proc. Natl. Acad. Sci. USA {\bf 103}, 19219 (2006).
\bibitem{CN4} P. Larimer and B. W. Strowbridge, J. Neurosci. {\bf 28}, 12212 (2008).
\bibitem{CN5} E. Bullmore and O. Sporns, Nat. Rev. Neurosci. {\bf 10}, 186 (2009).
\bibitem{CN6} O. Sporns, G. Tononi, and G. M. Edelman, Cereb. Cortex {\bf 10}, 127 (2000).
\bibitem{CN7} D. S. Bassett and E. Bullmore, The Neuroscientist {\bf 12}, 512 (2006).
\bibitem{SF1} P. Bonifazi, M. Goldin, M. A. Picardo, I. Jorquera, A. Cattani, G. Bianconi, A. Represa, Y. Ben-Ari, and R. Cossart, Science {\bf 326}, 1419 (2009).
\bibitem{SF2} C. Wiedemann, Nat. Rev. Neurosci. {\bf 11}, 74 (2010).
\bibitem{SF3} X. Li, G. Ouyang, A. Usami, Y. Ikegaya, and A. Sik, Biophys. J. {\bf 98}, 1733 (2010)
\bibitem{SF4} R. J. Morgan and I. Soltesz, Proc. Natl. Acad. Sci. USA {\bf 105}, 6179 (2008).
\bibitem{SF5} V. M. Egu{\'{i}}luz, D. R. Chialvo, G. A. Cecchi, M. Baliki, and A. V. Apkarian, Phys. Rev. Lett. {\bf 94}, 018102 (2005).
\bibitem{SF6} M. P. Young, Philos. Trans. R. Soc. {\bf 252}, 13 (1993).
\bibitem{SF7} M. P. Young, J. W. Scannell, G. A. Burns, and C. Blakemore, Rev. Neurosci. {\bf 5}, 227 (1994).
\bibitem{SF8} J. W. Scannell, C. Blakemore, and M. P. Young, J. Neurosci. {\bf 15}, 1463 (1995).
\bibitem{SF9} D. J. Felleman and D. C. Van Essen, Cereb. Cortex {\bf 1}, 1 (1991).
\bibitem{SF10} J. W. Scannell, G. A. P. C. Burns, C. C. Hilgetag, M. A. O'Neill, and M. P. Young, Cereb. Cortex {\bf 9}, 277 (1999).
\bibitem{SF11} O. Sporns, D. R. Chialvo, M. Kaiser, and C. C. Hilgetag, Trends Cogn. Sci. {\bf 8}, 418 (2004).
\bibitem{SF12} M. Kaiser, R. Martin, P. Andras, and M. P. Young, Eur. J. Neurosci. {\bf 25}, 3185 (2007).
\bibitem{BA1} A.-L. Barab\'{a}si and R. Albert, Science {\bf 286}, 509 (1999).
\bibitem{BA2} R. Albert and A.-L. Barab\'{a}si, Rev. Mod. Phys. {\bf 74}, 47 (2002).
\bibitem{HR1} J. L. Hindmarsh and R. M. Rose, Nature {\bf 296}, 162 (1982).
\bibitem{HR2} J. L. Hindmarsh and R. M. Rose, Proc. R. Soc. London, Ser. B {\bf 221}, 87 (1984).
\bibitem{HR3} R. M. Rose and J. L. Hindmarsh, Proc. R. Soc. London, Ser. B {\bf 225}, 161 (1985).
\bibitem{Longtin} A. Longtin, Phys. Rev. E {\bf 55}, 868 (1997).
\bibitem{GABA} N. Brunel and X.-J. Wang, J. Neurophysiol. {\bf 90}, 415 (2003).
\bibitem{SDE}  M. San Miguel and R. Toral, in {\it Instabilities and Nonequilibrium Structures VI}, edited by J. Martinez, R. Tiemann, and E. Tirapegui (Kluwer Academic Publisher, Dordrecht, 2000), pp. 35-130.
\bibitem{Kernel} H. Shimazaki and S. Shinomoto, J. Comput. Neurosci. {\bf 29}, 171 (2010).
\bibitem{Bely} I. Belykh and M. Hasler, Chaos {\bf 21}, 016106 (2011).
\bibitem{SWN} D.J. Watts and S.H. Strogatz, Nature {\bf 393}, 440 (1998).
\bibitem{Plant} R. E. Plant, J. Math. Biol. {\bf 11}, 15 (1981).

\end{thebibliography}
\end{document}